\documentclass[aps, prd, 10pt, twocolumn, english, longbibliography, superscriptaddress, breaklinks=true, showpacs=false,showkeys, nofootinbib]{revtex4-2}
\usepackage[T1]{fontenc}
\usepackage[latin9]{inputenc}
\usepackage{color}
\usepackage{babel}
\usepackage{amsmath}
\usepackage{amsthm}
\usepackage{amsfonts}
\usepackage{amssymb}

\usepackage{subfigure}
\usepackage{graphicx}
\usepackage{physics} 

\usepackage{booktabs}
\usepackage{array}[=2016-10-06]
\usepackage{tabularx}

\newcolumntype{L}[1]{%
  >{\raggedright\arraybackslash}p{#1}%
}

\newcolumntype{Y}{%
  >{\raggedright\arraybackslash}X%
}
\usepackage{hyperref}
 
\begin{document}

\title{On the quasiblack-hole limit of rotating charged fluids}

\author{Marcos L. W. Basso}
%\email{marcoslwbasso@hotmail.com}
\affiliation{Centro de Ci\^encias Naturais e Humanas, Universidade Federal do ABC, Avenida dos Estados 5001, Santo Andr\'e, S\~ao Paulo, 09210-580, Brazil}
\affiliation{Departamento de Matem\'atica Aplicada, Universidade Estadual de Campinas, 13083-859 Campinas, S\~ao Paulo, Brazil}
\author{Vilson T. Zanchin}
%\email{zanchin@ufabc.edu.br}
\affiliation{Centro de Ci\^encias Naturais e Humanas, Universidade Federal do ABC, Avenida dos Estados 5001, Santo Andr\'e, S\~ao Paulo, 09210-580, Brazil}

\begin{abstract}
We investigate the black hole limit of stationary, axisymmetric charged perfect fluids undergoing rigid or differential rotation, with and without pressure. Particular attention is given to Weyl-type configurations, in which the fluid redshift factor is functionally related to the generalized electromagnetic potential. In this limit, also called the quasiblack-hole (QBH) limit, the redshift factor vanishes throughout the fluid interior and the boundary becomes a quasihorizon. We focus on extremal QBH configurations with regular matter and electromagnetic fields and smooth matching to the exterior. Under suitable convergence assumptions, electromagnetic regularity and the approach to uniform rotation imply a constant generalized electromagnetic potential throughout the connected fluid interior, independently of the Weyl ansatz. With additional integrability conditions, the mass formula reduces to the Smarr relation for an extremal Kerr-Newman black hole. We also examine the compatibility of this limit with specific relations between the metric and electromagnetic potentials. For rigid rotation, these include charged dust satisfying a linear Weyl relation and fluids with pressure satisfying the Kloster-Das or Guilfoyle relations. The linear Kloster-Das subclass becomes pressureless in the limit, whereas the general Guilfoyle case allows nonzero pressure. The Islam ansatz imposes further restrictions that obstruct a regular limit when its coupling parameter, limiting potential, and limiting charge density are nonzero. For differential rotation, we analyze both configurations with an identically vanishing Lorentz-force term and a linear Weyl subclass whose regularity requires control of the angular-velocity gradients. Our results demonstrate that rotating Weyl-type systems admit extremal QBH limits in much the same way as their static counterparts. Moreover, 
these results extend analyses of rotating dust-fluid distributions and identify conditions under which more general rotating charged fluids are compatible with an extremal QBH limit.

\end{abstract}


\maketitle

\section{Introduction} 
\label{sec:introd}

The investigation of rotating fluid configurations in general relativity plays a crucial role in understanding the transition from compact objects to black holes. A remarkable property of certain equilibrium configurations of rotating fluid bodies is the existence of a black hole limit when a specific parameter threshold is reached. 
This phenomenon was first identified by Bardeen and Wagoner~\cite{Bardeen} in a numerical solution for a uniformly rotating dust disk. They observed that the exterior geometry of the disk approaches the extremal Kerr metric as a specific parameter reaches a critical value. 
Later, this same limit was found in several exact solutions describing relativistic rotating disks composed of dust~\cite{Neuge93, Neuge97, Ansorg00}, two counter-rotating streams of dust~\cite{Frauen, Ledvinka}, and a pressure-supported fluid~\cite{Bicak, Pichon}. Moreover, this limit was also observed numerically for rotating fluid configurations with toroidal topology~\cite{Ansorg, Fischer, Ames}. As described in Refs.~\cite{Meinel02, Meinel04}, the existence of such a limit implies a splitting of spacetime into two disjoint regions. From an exterior perspective, the extremal Kerr (or Kerr--Newman) metric emerges with its characteristic features, including a degenerate horizon and an infinitely long throat region, demonstrating that the black hole limit occurs independently of the fluid body's topology. From an interior perspective, the solution consists of a regular, non-asymptotically flat spacetime that exhibits the extremal Kerr near-horizon geometry at spatial infinity.  
The interior is filled with a stable, regular fluid or matter source. Because this spacetime describes an interior body rather than an isolated object in an empty universe, it does not flatten out into a void. Instead of flattening out, the interior spacetime transitions at its own spatial infinity into the exact near-horizon geometry of an extremal Kerr black hole.

Strictly speaking, a true black hole solution does not form upon reaching this ultra-compact limit. Therefore, as noted by Lemos and Zaslavskii~\cite{LemosZaslavskii2007}, these configurations are more accurately described as "quasiblack holes" (QBHs), meaning extremely compact systems on the threshold of genuine black hole formation. The term was originally coined by Lue and Weinberg~\cite{Lue1999, lue2}, and an explicit model within general relativity was subsequently investigated by Lemos and Weinberg in Ref.~\cite{Lemos:2003gx}. 
That work highlighted that such configurations had previously been identified by Bonnor~\cite{Bonnor:1972wi}, among others, as special limits of extremally charged stars. However, the general properties of QBHs were systematically studied a few years later by Lemos and Zaslavskii~\cite{LemosZaslavskii2007, Lemos2008, Lemos2008.1, Lemos2010, Lemos:2009wj, Lemos2020}. 

Since the pioneering work of Ref.~\cite{Lemos:2003gx}, several static solutions to the Einstein field equations have been shown to admit quasiblack-hole configurations in suitable limits (see, e.g., Refs.~\cite{kleber, Lemos:2006sj,lemoszanchin2008, Lemos:2010te,Arbanil:2013pua,Bronnikov:2013lha,Arbanil:2017huq,lemoszanchin2017,masa2023}). As these works demonstrate, the existence of a QBH limit is more likely in systems that also admit regular black hole configurations. This is particularly evident in Refs.~\cite{lemoszanchin2017,masa2023} for electrically charged configurations, and in Refs.~\cite{Bronnikov:2013lha,Yang:2021diz} for other kinds of systems. The connection between static regular black holes, quasiblack holes, and other horizonless nonrotating ultra-compact objects across a broad class of models is further supported by recent work in Ref.~\cite{Bronnikov:2024izh}.

While static quasiblack-hole models can be constructed by taking the appropriate limit of matter configurations typically used for regular horizonless compact objects or regular black holes, rotating counterparts cannot be obtained via a similar strategy, thereby necessitating alternative approaches. Attempts to extend static compact-object models, which admit a quasiblack-hole limit, to rotating configurations using standard solution-generating algorithms~\cite{Newman65, Gurses, Burinskii, Gondolo, Drake, Bambi, Mustapha} have frequently yielded Kerr-like geometries that fail to preserve the quasiblack-hole behavior of their static seeds. Although such procedures have successfully generated rotating regular black holes and other compact-object geometries from static seeds~\cite{Spallucci, Dymn15, Mazza, Franzin, Masa22, Maeda,Dymnikova:2024fce, BZ2024-paper1, Basso2025, Brustein, Dymnikova2025,Kim:2025sdj}, deriving a rotating metric in this manner does not inherently guarantee the existence of a corresponding family of fluid configurations that retain the quasiblack-hole limit. Specifically, these methods do not necessarily preserve the properties of the matter distribution and the physical relations that enable static configurations to achieve this limit. This motivates studying the quasiblack-hole limit directly from the governing equations of rotating charged fluids, without prescribing a particular equation of state, to identify the conditions under which such a limit remains compatible with fluid equilibrium and regularity.

As a matter of fact, following a strategy distinct from the metric-generation algorithms mentioned above, Meinel~\cite{Meinel06} demonstrated that increasing the compactness of a rigidly rotating fluid configuration yields solutions whose exterior geometry arbitrarily approaches that of an extremal Kerr black hole. This suggests that a rotating perfect fluid can, under appropriate conditions, mimic the exterior properties of a rotating black hole. For rigidly rotating disks of charged dust, the authors of Refs.~\cite{Meinel13, Meinel15} analyzed the quasiblack-hole limit toward an extremal Kerr--Newman black hole within the post-Newtonian expansion, finding that the limit exists only if the electromagnetic potential in the co-rotating frame is uniform across the disk. One of the aims of the present work is to extend these findings to rotating charged fluids with pressure, where the metric (redshift) function and the generalized electromagnetic potential satisfy the Weyl hypothesis~\cite{Weyl}.

Recently, a comprehensive analysis of rotating charged fluids with pressure satisfying Weyl's hypothesis was conducted in Ref.~\cite{basso2023}, extending several well-established theorems~\cite{Lemos:2009} from static Weyl-type systems to stationary and axisymmetric configurations. These static Weyl-type systems consist of charged fluid distributions that satisfy the Einstein-Maxwell field equations, where the electric potential $\phi$ and the metric potential $g_{tt}$ are related by $g_{tt} = g_{tt}(\phi)$. In the rotating case, it is the metric potential $g_{tt}$ in the co-rotating frame of the fluid, which measures the relative redshift of zero angular momentum photons emitted from the region of the fluid and received at infinity, that has a functional relation with the electromagnetic potential in the co-rotating frame. This relation between the metric and electromagnetic potentials imposes constraints on the fluid and electromagnetic field quantities. %which can be interpreted as an equation of state.

Here, we observe that the theorems established in Ref.~\cite{basso2023} can be used to explore the quasiblack-hole limit for rotating charged fluids, where the geometry approaches that of an extremal Kerr--Newman black hole, provided two conditions are met. Namely, the Smarr relation for an extremal Kerr--Newman black hole is satisfied in this limit, and the quasi-horizon becomes a null hypersurface fulfilling all the criteria characterizing the horizon of a stationary and axisymmetric extremal black hole~\cite{Meinel06, Meinel15}. Our results here indicate that rotating Weyl-type systems admit such a QBH limit in much the same way as their static counterparts, as described in Refs.~\cite{Bonnor:1972wi, kleber, Lemos:2006sj, lemoszanchin2008, Lemos:2010te}.

The present paper is organized as follows. In Sec.~\ref{sec:generalsetup}, we set the stage by considering rotating charged fluids in a stationary and axisymmetric spacetime, presenting the main equations and assumptions underlying our analysis. Important global quantities such as gravitational mass, angular momentum, and net electric charge are defined in terms of fundamental fields. Additionally, this section introduces the Smarr relation for a rotating charged fluid. Section~\ref{sec:qbhlimit} is devoted to revisiting the definition of quasiblack holes, discussing their main properties, and analyzing how the QBH limit is formally attained for rotating charged fluids. In Sec.~\ref{sec:rigid}, we investigate the QBH limit of Weyl-type rigidly rotating charged fluids, whereas in Sec.~\ref{sec:differential}, we analyze the corresponding limit for differentially rotating ones. Section~\ref{sec:conclusion} is devoted to summarizing and discussing the significance of the main results obtained. Finally, Appendix~\ref{sec:appendix} provides technical details utilized in the derivation of the Smarr relation.

\section{Rotating Einstein-Maxwell charged fluids}
\label{sec:generalsetup}

\subsection{Basic equations}
\label{sec:basiceqs}

To investigate the quasiblack-hole limit of rotating charged fluids within the framework of Einstein-Maxwell theory, we present in this section the fundamental equations governing such systems. We consider an electrically charged perfect fluid, so that the Einstein-Maxwell field equations may be written as
\begin{align}
    & G_{\mu \nu} =8\pi T_{\mu\nu} = 8\pi\left(E_{\mu \nu} +M_{\mu\nu}\right),\label{eq:Einst}\\
    &\nabla_{\nu} F^{\mu \nu} = 4 \pi J^{\mu}, \label{eq:Maxw}
\end{align}
where Greek indices range from $0$ to $3$. 

In Eq.~\eqref{eq:Einst}, $G_{\mu \nu}$ stands for the Einstein tensor, which is given in terms of the Ricci tensor $R_{\mu \nu}$ by the well-known relation $G_{\mu \nu}= R_{\mu \nu}- \frac{1}{2}g_{\mu \nu} R$, where $g_{\mu \nu}$ represents the metric tensor, and $R$ is the Ricci scalar. The energy-momentum tensor is decomposed as $T_{\mu \nu} = E_{\mu \nu} + M_{\mu \nu}$, where $ E_{\mu \nu}$ indicates the energy-momentum tensor due to the electromagnetic field alone, while $M_{\mu \nu}$ corresponds to the matter-energy tensor.

In Eq.~\eqref{eq:Maxw},  $\nabla_{\mu}$ represents the covariant derivative compatible with the four-dimensional Lorentzian metric, $F^{\mu\nu}$ stands for the Faraday-Maxwell field tensor, which can be expressed in terms of the gauge potential $A_\mu$ as $F_{\mu \nu} =  \nabla_{\nu} A_{\mu} - \nabla_{\mu} A_{\nu}$, and $J^\mu$ corresponds to the electric current density.

The energy-momentum tensor for the matter is given by
\begin{align}
    M_{\mu \nu} = \big(\rho_m + p\big)u_{\mu}u_{\nu} + p g_{\mu \nu}, \label{eq:fluid-emt}
\end{align}
with $\rho_m$ representing the energy density of the matter, $p$ denoting the fluid pressure, and $u^{\mu}$ being the four-velocity of the fluid, which satisfies the normalization condition $u^{\mu}u_{\mu} = -1$.

The electromagnetic contribution to the total energy-momentum tensor,  $E_{\mu \nu} $, is given by
\begin{align} 
    E_{\mu \nu} = \frac{1}{4\pi}\left( F_{\mu \alpha} F_{\nu}^{\ \alpha} - \frac{1}{4}g_{\mu \nu} F_{\alpha \beta}F^{\alpha \beta}\right). \label{eq:max-emt}
\end{align}

Furthermore, we assume a purely convective current density of the form
\begin{equation} 
J^{\mu} = \rho_e u^{\mu}, \label{eq:current}
\end{equation}
where $\rho_e$ represents the electric charge density.

For future reference, let us also define here the electric field $E_{\mu}$ and the magnetic field $B_{\mu}$, 
\begin{equation}
\begin{aligned}
    E_{\mu} &\equiv F_{\mu \nu} u^{\nu}  \label{eq:EBfields}\\
    B_{\mu} & \equiv - \frac{1}{2} \epsilon_{\mu \nu \alpha \beta} F^{\alpha \beta}  u^{\nu},
\end{aligned}
\end{equation}
with $ \epsilon_{\mu \nu \alpha \beta} $ being the Levi-Civita tensor.

It is also useful to deal with the electromagnetic energy density as measured by an observer whose four-velocity is the fluid velocity $u^\mu$, i.e., $\rho_{em} \equiv E_{\mu\nu} u^\mu u^\nu$. Using the definitions in Eq.~\eqref{eq:EBfields}, the electromagnetic energy density $\rho_{em}$ can be decomposed into its electric and magnetic components, i.e.,
\begin{align}
\rho_{em} = E_{\mu\nu} u^\mu u^\nu\equiv \rho_{el} + \rho_{mg} \label{eq:rhoem}
\end{align}
where 
\begin{align}
    & \rho_{el} \equiv \frac{1}{8 \pi} E_{\mu} E^{\mu}, \label{eq:rhoel0} \\
   & \rho_{mg} \equiv \frac{1}{8 \pi} B_{\mu} B^{\mu}, \label{eq:rhomag0} 
\end{align}
respectively. 

As we shall see, this decomposition will be useful to interpret the energy components as perceived by an observer comoving with the rotating charged fluid. 

The conservation equations 
\begin{align}
\nabla_{\nu} T^{\mu \nu} = 0, \label{eq:bianchi}\\
\nabla_{\mu} J^{\mu} = 0, \label{eq:conscharge}
\end{align}
are also useful for the present analysis. Combining these two equations and using the field equations \eqref{eq:Einst} and \eqref{eq:Maxw}, together with \eqref{eq:fluid-emt}, \eqref{eq:max-emt}, and \eqref{eq:current}, we get the following set of equations
\begin{align}
       & u^\nu \nabla_\nu \rho_m + (\rho_m + p) \nabla_\nu u^\nu = 0, \label{eq:continuity}\\
    & u^\nu \nabla_\nu \rho_e + \rho_e \nabla_\nu u^\nu=0. \label{eq:continuity1}\\
     & (\rho_m + p) u^{\nu} \nabla_\nu u^{\mu} + \nabla^{\mu}p = J_{\nu} F^{\mu \nu}, \label{eq:releuler}
\end{align}
Equation \eqref{eq:releuler} is the relativistic Euler equation, which results in the equilibrium equation for stationary systems. Equations \eqref{eq:continuity} and \eqref{eq:continuity1} are the continuity equations for matter and charge, respectively. 
From Eqs.~\eqref{eq:current}  and~\eqref{eq:EBfields}, we can see that the right-hand side of Eq.~\eqref{eq:releuler} is the Lorentz force perceived by an observer comoving with the fluid, once $J_{\nu} F^{\mu \nu} = \rho_e E^{\mu}$.

\subsection{Setting the stage: stationary and axisymmetric spacetimes}

The general class of spacetimes we are interested in here exhibits two fundamental symmetries, namely, stationarity and axisymmetry.  Moreover, we also assume that spacetime is circular~\cite{Wald}, which implies that, in the Weyl--Papapetrou coordinate system $(x^0, x^1, x^2, x^3) = (t, r, z, \varphi)$, the metric can be written in the form~\cite{Islam1978, Bonnor1980b} 
\begin{align}
    ds^2 = -f\, dt^2 + 2k\, dt d\varphi + l\, d \varphi^2 + e^{\mu}\left(dr^2 + dz^2\right), \label{eq:metric3}
\end{align}
where the metric functions $f$, $k$, $l$, and $\mu$ depend solely on $r$ and $z$. 
This geometry admits a timelike Killing vector field $\xi^{\mu} = \delta^{\mu}_{\ t}$, normalized such that $\xi_{\mu}\xi^{\mu} = -1$ at spatial infinity, and a spacelike Killing vector field $\eta ^{\mu}= \delta^{\mu}_{\ \phi}$, whose orbits are closed curves and which vanishes on the symmetry axis.%normalized such that $\eta ^{\mu }\eta_\mu=1$.

Since the spacetime exhibits circular isometry along $\eta^\mu$, the four-velocity of the fluid describes a purely circular motion around the axis of rotation, which allows us to write the four-velocity as
\begin{align}
       u^{\mu} =&\, {\cal F}^{-1}\left(\xi^{\mu} + \Omega \eta^{\mu}\right),\label{eq:4velocity} 
\end{align}
with
\begin{equation} 
\mathcal{F}^2  \equiv  f - 2k\Omega - l \Omega^2, \label{eq:calF}
\end{equation}
and where
\begin{align}
    \Omega \equiv \frac{u^{\varphi}}{u^t}  = \frac{d \varphi}{dt}
\end{align}
is the angular velocity of the fluid. The function $\mathcal{F}$ is referred to as the redshift factor~\cite{thorne}, as it is connected to the relative redshift $\mathcal{Z}$ of photons with zero angular momentum emitted from the fluid region and received at infinity, given by~\cite{Meinel06}
\begin{equation}
    \mathcal{Z} = {\cal F}^{-1} - 1.\label{eq:redshift}
\end{equation}

Assuming that the electromagnetic field also respects the spacetime symmetries, we can express the gauge potential as
\begin{align}
A_{\mu} =&\, \phi \,\delta_{\mu}^{\ t} + \psi\, \delta_{\mu}^{\ \varphi}, \label{eq:gaugepot}
\end{align}
where $\phi=\phi(r,z)$ and $\psi=\psi(r,z)$ are the electric and magnetic potentials, respectively. After this, the electric and magnetic fields, as defined in Eq.~\eqref{eq:EBfields}, are given by
\begin{align}
    & E_{\mu}  = -\delta_{\mu}^{\ j} \mathcal{F}^{-1} \big(\nabla_j \Phi - \psi \nabla_j \Omega\big), \label{eq:rotE} \\
    & B_{\mu} = - \delta_{\mu}^i \epsilon_{t i j \varphi}\frac{\mathcal{F}^{-1}}{D^2}\big(K \nabla^j \Phi + F \nabla^j \psi - K \psi \nabla^j \Omega\big),\label{eq:rotB}  
\end{align}
respectively.

To simplify notation, in the last equations we have introduced the following definitions,
\begin{align}
& \Phi = \phi + \Omega \psi, \label{eq:Phi}\\
& F = \mathcal{F}^2=f - 2k\Omega - l \Omega^2, \label{eq:F} \\
& D^2 =fl + k^2, \label{eq:D2} \\
& K = k + l \Omega. \label{eq:K}    
\end{align}
Furthermore, $\nabla_i$ denotes the covariant derivative associated with the induced metric $g_{ij}$ on the $r-z$ submanifold. Considering the full metric \eqref{eq:metric3}, we find $g_{ij}=e^{\mu}\delta_{ij}$, where the indices $i,\,j,\, \ldots$ run from $1$ to $2$. Since the relevant scalar quantities are independent of the coordinates $t$ and $\varphi$, the four-dimensional covariant derivative $\nabla_\mu$ reduces to $\nabla_i$. 

The field $\Phi$ can be interpreted as an effective electromagnetic potential.
The metric function $D$ also has a simple interpretation, at least when dealing with pressureless fluids. In fact, from the Einstein equations one obtains
\begin{align}
R^t_{\ t} + R^{\varphi}_{\ \varphi}
= -D^{-1} e^{- \mu}\left(\partial_r^2 D + \partial_z^2 D \right)
= -16\pi p.
\end{align}
Hence, in the dust case ($p=0$), the function $D$ satisfies the Laplace equation and is therefore harmonic. Without loss of generality, one may choose the canonical Weyl coordinates such that $D^2 = r^2$. For fluids with nonvanishing pressure, however, $D$ is no longer harmonic, and this canonical choice is, in general, not available (See Ref.~\cite{Islam1985} for further details). In addition, one can show that
\begin{align}
    D^2 = FL + K^2,
\end{align}
where $L=l$. As we will discuss below (see Sec.~\ref{sec:corotating}), the quantities $F$, $L$, and $K$ also arise naturally in the corotating frame for rigidly rotating fluids.

Then, using the definitions \eqref{eq:rhoel0} and \eqref{eq:rhomag0}, the electric and magnetic energy densities, as measured by a local observer co-rotating with the fluid, are given by
\begin{align}
 & \rho_{el} = \frac{1}{8 \pi F} \big(\nabla \Phi - \psi \nabla \Omega\big)^2, \label{eq:grhoel1} \\
 & \rho_{mg} = \frac{1}{8\pi D^2 F}\big(K \nabla \Phi + F \nabla \psi - K \psi \nabla \Omega\big)^2, \label{eq:grhomg2}
\end{align}
respectively. Here, we use the notation $(\nabla Q)^2 \equiv (\nabla Q)\cdot(\nabla Q)\equiv \nabla_i Q \nabla^i Q$, where $Q$ denotes any scalar field.

It is worth mentioning that, for differential rotation, a global co-rotating frame with the fluid does not exist. Instead, co-rotation is only defined locally, that is, within each layer of the fluid where $\Omega(r,z)$ remains constant. 

Now we deal with the conservation equations \eqref{eq:bianchi} and \eqref{eq:conscharge}, or, equivalently, with Eq.~\eqref{eq:continuity}, \eqref{eq:continuity1}, and \eqref{eq:releuler}.
First we note that, due to the symmetries of spacetime and circularity, Eqs.~\eqref{eq:continuity} and~\eqref{eq:continuity1} are trivially satisfied. On the other hand, from the relativistic Euler equation~\eqref{eq:releuler}, we obtain the equilibrium equation
\begin{equation}
\begin{aligned}
&   \left(\rho_m+p\right)\nabla_i\mathcal{F} +\rho_e\left(\nabla_i\Phi - \psi\nabla_i\Omega\right)  \\
&\; +\left[\left(\rho_m +p\right)\mathcal{F}^{-1} K\right]\nabla_i\Omega + \mathcal{F}\nabla_i p=0, \label{eq:relequi}
\end{aligned}
\end{equation}
with the index $i$ running from $1$ to $2$ and $ K = k +\Omega\, l$ is the relativistic centrifugal potential.

Although the physical interpretation of each term in Eq.~\eqref{eq:releuler} is well-known, it is worth recalling it at this point. The equilibrium equation indicates the balance of different forces acting  upon a fluid element. We identify the gravitational force $\left(\rho_m+p\right)\nabla_i\mathcal{F}$, the Lorentz force $\rho_e(\nabla_i\Phi - \psi \nabla_i \Omega) 
=\rho_e\left(\nabla_i\phi + \Omega \nabla_i \psi\right)$, the centrifugal force $\left(\rho_m +p\right)\mathcal{F}^{-1} K\nabla_i\Omega$, and the pressure force $\mathcal{F}\,\nabla_i p$. 
It is also worth noting that the quantity $\big(\rho_m + p\big){\cal F}^{-1}K = \big(\rho_m + p\big)u_{\varphi}$ in Eq.~\eqref{eq:relequi} represents the angular momentum density of the fluid. Similarly, the term $\rho_e \psi$ can be understood as representing the angular momentum density associated with the electromagnetic field. %the combination
Throughout this work, the term "Lorentz force" refers specifically to the combination
\begin{equation} 
\rho_e\left(\nabla_i\Phi-\psi\nabla_i\Omega\right) =-\mathcal F\, \rho_e E_i, \label{eq:comovingLorentzforce} 
\end{equation}
that enters the equilibrium equation. Consequently, when considering configurations with a vanishing Lorentz force in the subsequent sections, we mean precisely that the quantity $\nabla_i\Phi-\psi\nabla_i\Omega$ vanishes, rather than the comoving electric field going to zero. This distinction is crucial in the QBH limit, where it is $\mathcal{F}$ that vanishes, rather than $E_{i}$.

Other important quantities for the present work are the gravitational mass $M$, the angular momentum $J$, and the electric charge $Q$ of the rotating fluid distribution. They are defined by~\cite{Wald}
\begin{align}
    & M = 2 \int_{\Sigma_t}\left(T_{\mu \nu} - \frac{1}{2}g_{\mu \nu} T \right) \xi^{\mu} n^{\nu} d V , \label{eq:mass}\\
    & J = -\int_{\Sigma_t}T_{\mu \nu} \eta^{\mu} n^{\nu} d V , \label{eq:angmom} \\
    & Q = - \int_{\Sigma_t} J_{\mu} n^{\mu} d V,
\end{align}
respectively. Here, $\Sigma_t$ is a spacelike hypersurface defined by $t = \text{constant}$, $dV$ is the volume element in $\Sigma_t$, $dV = \sqrt{h} d^3x$, $n^{\mu}$ is future pointing unit normal to $\Sigma_t$ such that $d V^{\mu} = - n^{\mu} dV$ and $d V^{\mu} n_{\mu} = dV = \sqrt{h}\, d^3x$, with $h$ being the determinant of the induced metric in $\Sigma_t$.  Eqs.~\eqref{eq:mass} and~\eqref{eq:angmom} can be further combined into
\begin{align}
M - 2 \Omega J & = 2\! \int_{\Sigma_t}\!\!\left(T_{\mu \nu} - \frac{1}{2}g_{\mu \nu} T \right) \chi^{\mu} n^{\nu} d V \label{eq:mJ} \\ 
& + 2 \Omega\! \int_{\Sigma_t}\!T_{\mu \nu} \eta^{\mu} n^{\nu} d V - 2 \!\int_{\Sigma_t}\!\! \Omega \,T_{\mu \nu} \eta^{\mu} n^{\nu} d V, \nonumber
\end{align}
where
\begin{align}
\chi^{\mu} = \xi^{\mu} + \Omega\, \eta^{\mu} = {\cal F} u^\mu \label{eq:chi}  
\end{align}
defines a linear combination between the Killing vectors $\xi^{\mu}$ and $\eta^{\mu}$. If the charged fluid rotates rigidly, i.e., $\Omega = \text{constant}$, then $\chi^{\mu}$ is also a Killing vector. 
Also note that, in the case of rigid rotation, the last two terms on the right-hand side of Eq.~\eqref{eq:mJ} cancel out. On the other hand, if the charged fluid undergoes differential rotation, i.e., $\Omega = \Omega(r,z)$, then $\chi^{\mu}$ is not a Killing vector, and the last two terms on the right-hand side of Eq.~\eqref{eq:mJ} do not cancel out in general.

A straightforward calculation shows that Eq.~\eqref{eq:mJ} becomes
\begin{equation}
\begin{aligned}
 M & -  2 \Omega J - \Phi\, Q 
=  -\int_{V_m}\left(\rho_m + 3p\right)\mathcal{F}\, u_{\mu} n^{\mu} d V \\ 
 &  -\int_{V_m} \Phi\, \rho_e u_{\mu} n^{\mu} d V + \Phi\int_{V_m} \rho_e u_{\mu}n^{\mu} dV 
 \\
 & -\frac{1}{2\pi} \int_{V_m} \psi F^{\mu \nu} n_\mu \left(\nabla_{\nu} \Omega\right)\,  dV \label{eq:generalsmarr}  \\ 
 & - \frac{1}{4\pi} \int_{V_m} A_{\nu} F^{\mu \nu} \left(\nabla_{\mu} \Omega\right)\eta^{\sigma} n_\sigma dV \\ 
 & + 2 \Omega\int_{\Sigma_t}T_{\mu \nu} \eta^{\mu} n^{\nu} d V - 2 \int_{\Sigma_t} \Omega\, T_{\mu \nu}  \eta^{\mu} n^{\nu} d V, 
\end{aligned}
\end{equation}
with $V_m \subset \Sigma_t$ being the spatial region where there is charged matter. In Appendix~\ref{sec:appendix}, we describe the main steps of the derivation of Eq.~\eqref{eq:generalsmarr}. We have rearranged this equation by adding and subtracting terms involving $\Phi Q$ and $\Omega J$ to isolate the combination entering the extremal Smarr relation on the left-hand side. As shown below, this form is particularly convenient for taking the QBH limit.

It is worth mentioning that Eq.~\eqref{eq:generalsmarr} with $\rho_e = 0$ and $\nabla_{\mu} \Omega  = 0$ was used in Ref.~\cite{Meinel06} to probe the QBH limit of uncharged rigidly rotating fluid bodies in equilibrium. In addition, Eq.~\eqref{eq:generalsmarr} with $p = 0$ and $\nabla_{\mu} \Omega  = 0$ was used in Ref.~\cite{Meinel15} to probe the QBH limit of a disk composed of rigidly rotating charged dust. Hence, Eq.~\eqref{eq:generalsmarr} generalizes the equations obtained in Refs.~\cite{Meinel06, Meinel15} and will play a major role in the present work, in order to probe the QBH limit.

\section{ Defining the quasiblack-hole limit of rotating charged fluid distributions}
\label{sec:qbhlimit}
We review here the definition of a rotating QBH following closely Ref.~\cite{Lemos:2009wj} and discuss how the QBH limit is attained for the rotating charged fluid considered in the previous section. 

To this end, let us rewrite the metric given by Eq.~\eqref{eq:metric3} as
\begin{align}
ds^2 = -N\, dt^2 + l\left(d\varphi - \omega dt \right)^2
+ e^{\mu}\left(dr^2 + dz^2\right), \label{eq:metric4}
\end{align}
where 
\begin{align}
     N = f + l \omega^2 ,  \ \ \ \omega = -\frac{k }{l}, \label{eq:shift}    
\end{align}
with $N$ being the lapse function and $\omega$ being the angular velocity acquired by an observer freely falling from infinity. In terms of the redshift function ${\cal F}^2$, the lapse function $N$ reads
\begin{equation}
N=  \mathcal{F}^2 + l(\Omega- \omega)^2. \label{eq:lapse}
\end{equation}
Furthermore, in terms of the lapse and redshift functions the metric function $D^2$ \eqref{eq:D2} assumes the form
\begin{equation}
D^2 = l N = l\left(  \mathcal{F}^2 + l(\Omega- \omega)^2 \right) \label{eq:D2b}
%\dfrac{N}{l} = \dfrac{\mathcal{F}^2}{ l}  + (\Omega- \omega)^2. \label{eq:D2b}
\end{equation}

Let us consider a spacetime configuration that depends on a parameter $\varepsilon$, and satisfies the following conditions:\vskip .1cm
\noindent 
(i) For small but nonzero values of $\varepsilon$, the metric is regular throughout the spacetime, with a nonvanishing lapse function $N>0$;\vskip .1cm
\noindent 
(ii) If $\varepsilon$ is the maximum value of the lapse function at the boundary, $N_B$, then in the limit $\varepsilon \to 0$, the lapse function satisfies $N \leq N_B \to 0$ everywhere within the inner region; \vskip .1cm
\noindent 
(iii) The Kretschmann scalar remains finite at the horizon (or quasi-horizon) in the limit $\varepsilon \to 0$. This latter property implies that the spatial area of the quasi-horizon reaches a minimum value;
\vskip .1cm
\noindent 
(iv) In the limiting case under consideration, the angular velocity $\omega$ approaches a constant value $\omega_h$ throughout the inner region.

Condition (i) introduces an infinitesimal parameter $\varepsilon$, which is employed in conditions (ii) and (iii). 
Additionally, condition (iv) introduces a second infinitesimal parameter. Namely, we may write $\left|\omega - \omega_h\right| \leq \delta$, with $\delta$ being a small nonnegative parameter.  
The QBH limit is obtained by taking $\varepsilon \to 0$ and $\delta\to 0$.  Notice that the condition $\omega\to\omega_h$ is a necessary condition, otherwise differential rotation within the interior region would distinguish the fluid distribution in the limit  $\varepsilon\to 0$ from a true black hole.  

In the present case, the limit $N\to\varepsilon $ implies $\mathcal{F}^2 \to \varepsilon$, which can be seen using relation \eqref{eq:lapse}. In fact, such relation  implies that $\mathcal{F}^2 + l(\Omega- \omega)^2\to \varepsilon$ and,  since we are interested in situations where $\mathcal{F}^2 >0$ with nonvanishing $l>0$, it follows that both quantities $\mathcal{F}^2$ and  $(\Omega- \omega)^2$  must approach $\varepsilon\to 0$ at the same rate, i.e., we must have $\mathcal{F}^2\to \varepsilon$ and  $(\Omega- \omega)^2\to \varepsilon$.  These results can be rewritten as 

\begin{equation}
N \to \varepsilon \to 0
\quad \Longrightarrow \quad
\left\{
\begin{aligned}
& \mathcal{F}^2 \to \varepsilon \to 0,\\
&\left|\Omega-\omega\right| \to \sqrt{\varepsilon} \to 0.
\end{aligned}
\right.
\label{eq:limit1}
\end{equation}
Notice that condition \eqref{eq:limit1} yields a nontrivial rotating limit only if $\Omega$ and $\omega$ have the same sign. Indeed, if they
had opposite signs, one would have $|\Omega-\omega|=|\Omega|+|\omega|,$ and condition \eqref{eq:limit1} would imply $\Omega\to 0, \
\omega\to 0,$ corresponding to a static limit. Therefore, in the following we restrict
ourselves to configurations for which $\Omega$ and $\omega$ have the same
sign. Without loss of generality, we assume $\Omega>0$ and $\omega>0$. Now, by the triangle inequality,
\begin{equation}
   \left|\Omega-\omega_h\right| = \left|\Omega- \omega + \omega -\omega_h\right| 
     \le |\Omega- \omega| + | \omega - \omega_h|.
  \label{eq:Omegalimit}
\end{equation} 
After relation \eqref{eq:Omegalimit}, in the QBH limit one has 
\begin{equation}
     \left|\Omega-\omega_h\right| \leq  \sqrt{\varepsilon} + \delta. 
\label{eq:omegaslimit}
\end{equation} 
Therefore, using condition (iv) together with condition~\eqref{eq:limit1}, specifically, $|\Omega- \omega| \to \sqrt{\varepsilon} \to 0$ and $| \omega - \omega_h| \to \delta \to 0 $, it follows that
\begin{align}
    \left|\Omega-\omega_h\right| \to 0. 
\end{align}

Additionally, relation \eqref{eq:limit1} implies also that the metric potential $K$, defined in Eq.~\eqref{eq:K} also vanishes in the QBH limit. In fact, from Eqs.~\eqref{eq:K}  and \eqref{eq:shift} we may write $K= l\left(\Omega - \omega\right) $ and then, using Eq.~\eqref{eq:limit1} it follows 
\begin{equation}
    K\to \sqrt{\varepsilon}\to 0. \label{eq:Klimit}
\end{equation}

The last important metric function we mention at this point is $D^2$. After relations \eqref{eq:D2b} and \eqref{eq:limit1}, it follows 
\begin{equation}
    D^2 \to \varepsilon. \label{eq:limitD2}
\end{equation}

In the case of differential rotation, we assume that the fluid angular velocity $\Omega$ approaches the angular velocity of QBH $\omega_h$ in one of the following two ways:
\begin{itemize}
    \item[(a)] $\Omega \to \omega \to \omega_h$ through a sequence of functions $\Omega_n$ and $\omega_n$ such that $0 \le \Omega_1 \le \Omega_2 \le \cdots \le \infty$ and $0 \le \omega_1 \le \omega_2 \le \cdots \le \infty$, with $\lim_{n \to \infty} \Omega_n = \lim_{n \to \infty} \omega_n = \omega_h$ throughout the entire fluid region;

    \item[(b)] $\Omega \to \omega \to \omega_h$ through a sequence of functions $\Omega_n$ and $\omega_n$ such that $\lim_{n \to \infty} \Omega_n = \lim_{n \to \infty} \omega_n = \omega_h$ almost everywhere in the fluid region, and $|\Omega_n| \le f_{\Omega}$, $|\omega_n| \le f_{\omega}$ throughout the region, where $f_{\Omega}$ and $f_{\omega}$ are integrable functions.
\end{itemize}

Condition (a) ensures that the \textit{monotone convergence theorem} can be applied, while condition (b) allows for the use of the \textit{Lebesgue-dominated convergence theorem}. In both cases, it is possible to interchange the limit and the integration sign under sufficiently general assumptions and in any measure space~\cite{Rudin}. Let us notice that condition (a) is more natural in the present context, as the  definition of a QBH requires that $\omega \to \omega_h$ throughout the entire interior region, and therefore $\Omega \to \omega \to \omega_h$ also in that region.

It is worth noting that condition (b) does not require the sequence $\Omega_n$ to be monotonic, it suffices that it converges almost everywhere and is bounded by an integrable function. In our specific context, applying condition (b) requires the stronger assumption that $\Omega_n$ converges everywhere. For this reason, condition (a) is adopted as the default, even though, in principle, the more restrictive version of condition (b) could be employed instead. As demonstrated below, condition (a) is essential for analyzing the QBH limit within the framework of a charged fluid undergoing differential rotation.

Nevertheless, the monotonic convergence of functions $\Omega_n$ says nothing about the behavior of their derivatives. Therefore, it is necessary to assume additional convergence hypotheses to ensure that $\nabla_\mu \Omega \to \nabla_\mu \omega_h$ when $\Omega \to \omega_h$. More precisely, we assume that the sequences $\Omega_n$ and $\omega_n$ are differentiable in the fluid region and converge pointwise (in addition to monotonically) to $\omega_h$. Furthermore, we assume that the derivatives $\nabla_\mu \Omega_n$ and $\nabla_\mu \omega_n$ converge uniformly throughout the fluid region. Under these conditions, we can state that $\lim_{n \to \infty} \nabla_\mu \Omega_n = \lim_{n \to \infty} \nabla_\mu \omega_n = \nabla_\mu \omega_h$. As a consequence, the gradient $\nabla_i \Omega$ approaches the QBH limit at the same rate as the difference $\left|\Omega-\omega_h\right|$ itself, cf. Eq.~\eqref{eq:omegaslimit},
\begin{equation}
 \left|\nabla_i \Omega\right| \to \sqrt{\varepsilon}+\delta. \label{eq:gradOmega}
\end{equation}
In other words, the limit and differentiation operators commute (for a detailed discussion on these subjects, see~\cite{Rudin}).
We shall assume analogous convergence properties for the metric functions $\mathcal{F}$, $F$, $K$, $\omega$, and also for the electromagnetic potentials $\phi$ and $\psi$ throughout the manuscript.

Table \ref{tab:qbh-metric-limit} summarizes the asymptotic behavior of the metric potentials in the QBH limit, while Table \ref{tab:qbh-fluid-limit} summarizes the asymptotic behavior of the fluid quantities and the potentials in the QBH limit.
Notice that the symbol $\nabla$ indicates the gradient of the respective function. 

The above analysis depicted the asymptotic behavior of the metric functions and of the fluid angular velocity $\Omega$. It is also useful to depict the asymptotic behavior of the other fluid quantities and of the electromagnetic potentials and energy density. %

We assume that the matter variables remain finite and well-behaved in the QBH limit, and we require the electromagnetic energy densities to remain finite throughout the fluid region. From Eq.~\eqref{eq:grhoel1}, given that $F \to \varepsilon$, the finiteness of the electric energy density demands that
\begin{align}
  \nabla_i\Phi-\psi\nabla_i\Omega \to \sqrt{\varepsilon}, \label{eq:lorentzforce}
\end{align}
Equation~\eqref{eq:lorentzforce} should be interpreted as a regularity condition: the combination $\nabla_i\Phi-\psi\nabla_i\Omega$ must vanish quickly enough to ensure that its ratio to $\sqrt{\varepsilon}$ remains bounded. 
This does not imply that the combination necessarily scales as $\sqrt{\varepsilon}$. More specifically, it may vanish faster or even be identically zero. Furthermore, Eq.~\eqref{eq:lorentzforce} does not constrain the asymptotic behavior of $\nabla_i\Phi$ and $\nabla_i\Omega$ individually. Their relative behavior is instead determined by the additional assumptions that define each configuration class examined below.

For rigidly rotating fluid configurations, $\nabla_i\Omega$ vanishes identically. In this case, Eq.~\eqref{eq:lorentzforce} reduces to $\nabla_i\Phi \to \sqrt{\varepsilon}$, which is also the behavior obtained for the Weyl-type relations considered in Sec.~\ref{sec:rigid}.

For the differentially rotating fluid configurations with a vanishing Lorentz force considered in Sec.~\ref{sec:VB}, a stronger condition holds: $\nabla_i\Phi-\psi\nabla_i\Omega=0$ identically throughout the fluid region. Therefore, Eq.~\eqref{eq:lorentzforce} is automatically satisfied, independently of the individual scaling of the two terms. The vanishing Lorentz force condition instead relates their gradients according to $\nabla_i\Phi=\psi\nabla_i\Omega$. Since $\psi$ remains finite according to the convergence assumptions introduced above, one has $\nabla_i\Phi \to \sqrt{\varepsilon}+\delta$.

The situation is different for the Weyl-type differentially rotating configurations considered in Sec.~\ref{sec:VC}. Here, no vanishing Lorentz force condition is imposed from the outset, and $\nabla_i\Phi$ and $\nabla_i\Omega$ are not generally related. Instead, a linear Weyl relation between $\mathcal F$ and $\Phi$ is assumed, which implies that $\nabla_i\Phi \to \sqrt{\varepsilon}$, whereas the convergence assumptions for the angular velocity give Eq.~\eqref{eq:gradOmega}. The relative scaling between $\delta$ and $\varepsilon$ must then be chosen so that the regularity condition~\eqref{eq:lorentzforce} remains satisfied, as it will be discussed explicitly in Sec.~\ref{sec:VC}.
 
A similar regularity condition follows from the magnetic energy density. Equation~\eqref{eq:grhomg2} may be conveniently rewritten as
\begin{align}
    \rho_{mg} = \frac{1}{8\pi D^2 F} \big[ K\left(\nabla\Phi-\psi\nabla\Omega\right) +F\nabla\psi \big]^2 . \label{eq:grrhomg22}
\end{align}
Since $D^2 \to \varepsilon$, $F \to \varepsilon$, $K \to \sqrt{\varepsilon}$, and the regularity of the electric energy density requires $\nabla\Phi - \psi\nabla\Omega \to \sqrt{\varepsilon}$, the first term inside the brackets in Eq.~\eqref{eq:grrhomg22} is already $\mathcal{O}(\varepsilon)$. Consequently, the finiteness of $\rho_{\rm mg}$ is guaranteed provided that $\nabla_i\psi \sim \mathcal{O}(1)$, i.e., the gradient of the magnetic potential remains finite in the QBH limit. We further assume that $\psi$ itself remains finite throughout the fluid region.

\begin{table}[t]
\begin{tabular}{|c|c|c|c|c|c|c|c|c|c|}
\hline
  Function &   $\ {\cal F}\ $    & $\ \nabla{\cal F} \ $   & $\ F\ $ & $\ \nabla F\ $ & $\ K\ $ & $\ \nabla K\ $ & $\ D^2\ $ & $\ \omega\ $ &$\ \nabla \omega\ $\\
    \hline 
Behavior &   $\sqrt{\varepsilon}$   &  $\sqrt{\varepsilon}$   & $\varepsilon$ &  $\varepsilon$ &  $\sqrt{\varepsilon}$   &  $\sqrt{\varepsilon}$ &  $\varepsilon $   &  $\ \omega_h\ $ &  $\delta$ \\
   \hline
\end{tabular}
\caption{The asymptotic behavior of the metric quantities in the QBH limit.}
\label{tab:qbh-metric-limit}
\end{table}

\begin{table}[t]
    \centering
    \renewcommand{\arraystretch}{1.3}
    \begin{tabular}{|l|c|c|c|c|}
        \hline
        Configuration
        & $\Omega$
        & $\nabla\Omega$
        & $\Phi$
        & $\nabla\Phi$
        \\
        \hline
        Rigid rotation
        & $\omega_h$
        & $ 0$
        & $\ \Phi_h $
        & $\sqrt{\varepsilon}$
        \\
        \hline
        Differential rotation: Sec.~V.B
        & $\ \omega_h\ $
        & $\ \sqrt{\varepsilon}+\delta\ $
        & $\ \Phi_h$
        & $\ \sqrt{\varepsilon}+\delta\ $
        \\
        \hline
        Differential rotation: Sec.~V.C
        & $\omega_h$
        & $ \sqrt{\varepsilon}+\delta$
        & $\ \Phi_h$
        & $\sqrt{\varepsilon}$
        \\
        \hline
    \end{tabular}
    \caption{The asymptotic behavior of the fluid angular velocity and electromagnetic potential in the QBH limit.}
    \label{tab:qbh-fluid-limit}
\end{table}

Before proceeding to analyze specific Weyl-type relations, it is useful to extract a general consequence of the equilibrium equation~\eqref{eq:relequi} in the QBH limit.

For rigidly rotating fluids, the angular velocity is constant throughout the fluid by definition, and therefore the equilibrium equation~\eqref{eq:relequi} yields
\begin{equation}
\left(\rho_m+p\right)\nabla_i\mathcal{F} +\rho_e\nabla_i\Phi + \mathcal{F}\,\nabla_i p=0. \label{eq:relequi-rigid}
\end{equation}

For differentially rotating fluids, on the other hand, the assumptions introduced above imply that the angular velocity approaches a constant value throughout the interior region ($\Omega\rightarrow\omega_h$), so that $\nabla_i\Omega\rightarrow0$ there. In fact, $\nabla_i\Omega$ vanishes at least as fast as $\sqrt{\varepsilon}$ in the QBH limit.
Therefore, provided that the coefficient $(\rho_m+p)\mathcal{F}^{-1}K-\rho_e\psi$ in Eq.~\eqref{eq:relequi} remains finite, the centrifugal contribution to the equilibrium equation vanishes identically in the QBH limit. Since both metric potentials $\mathcal{F}$ and $K$ approach zero at the same rate, the ratio $\mathcal{F}^{-1}K$ is finite, and the finiteness of the full coefficient is guaranteed by assuming that the functions $\rho_m+p$ and $\rho_e\psi$ take on finite values in the region of interest.
In this scenario, the equilibrium equation~\eqref{eq:relequi} reduces to the same form as that for rigidly rotating systems, cf. Eq. \eqref{eq:relequi-rigid}. 

Based on the results of the last two paragraphs, the QBH limit of the equilibrium relation \eqref{eq:relequi} for differentially rotating fluids is equivalent to that for rigidly rotating fluids. Moreover, since the redshift function satisfies $\mathcal{F} \to \sqrt{\varepsilon} \to 0$ throughout the interior region, its gradient also vanishes ($\nabla_i\mathcal{F} \to 0$). Assuming that the matter variables $\rho_m$, $p$, and $\nabla_i p$ remain finite in this region, both the gravitational term $(\rho_m+p)\nabla_i \mathcal{F}$ and the pressure term $\mathcal{F} \nabla_i p$ vanish in the QBH limit. Consequently, Eq.~\eqref{eq:relequi} reduces to $\rho_e\nabla_i\Phi=0$. Hence, for any charged fluid configuration with a finite, nonvanishing charge density $\rho_e$, we have $\nabla_i\Phi=0$, which implies that the electromagnetic potential becomes constant throughout the interior region (i.e., $\Phi \to \Phi_h$). In other words, the Lorentz force vanishes as a direct consequence of the QBH limit itself, independent of the particular Weyl-type relation satisfied by the fluid, as will be demonstrated in the subsequent sections.

Finally, we emphasize that the QBH limits considered throughout this work correspond to extremal QBHs, whose exterior geometry approaches that of an extremal Kerr--Newman black hole. This restriction is closely related to the regularity assumptions underlying our analysis. In addition to requiring the relevant matter and electromagnetic quantities to remain finite in the QBH limit, we restrict our focus to configurations where the matching between the interior and exterior regions is smooth, avoiding boundary layers or divergent surface stresses at the quasi-horizon. As demonstrated in previous studies of QBHs (Refs.~\cite{Lemos2008.1, Lemos:2009wj, Lemos2010}), nonextremal QBH limits generically develop divergent stresses at the boundary, whereas the extremal case allows the limiting configuration to remain regular. It is therefore natural, within the class of regular configurations investigated here, to assume a smooth interior-exterior matching and focus on the extremal limit (see Ref.~\cite{Lemos:2009wj} for the explicit matching procedure). This choice is also consistent with the distinctive geometric properties of extremal Kerr--Newman spacetime. In particular, its degenerate horizon and infinitely extended throat region allow the QBH limit to be defined independently of the topology of the fluid body~\cite{Meinel06}. From the exterior perspective, the horizon of an extremal Kerr--Newman black hole lies at an infinite proper distance from any point in the exterior region~\cite{Horowitz}, even though both the horizon and the limiting throat are located at $r_{\rm BH}=M$ in Boyer--Lindquist coordinates. For a more detailed discussion of this limiting geometry, see Refs.~\cite{Meinel02, Meinel04}.

\section{Quasiblack-hole limit of Weyl-type rigidly rotating charged fluids}
\label{sec:rigid}

\subsection{The basic equations in co-rotating coordinates}
\label{sec:corotating}

Let us now assume that the fluid is initially in rigid rotation, and then the QBH limit is taken. In this case,  $\Omega = \text{constant}$ and the linear combination between the Killing vectors in Eq.~\eqref{eq:chi} also defines a Killing vector field such that $\chi_{\mu}\chi^{\mu}= -\mathcal{F}^2$. Consequently, when the QBH limit is taken, $\chi^{\mu}$ becomes a null Killing vector everywhere inside the inner region. In particular, the quasi-horizon becomes a null hypersurface and fulfills all the criteria that characterize the horizon of a stationary and axisymmetric extremal black hole, with $\omega_h$ representing its angular velocity, as discussed in Ref.~\cite{Meinel06, Meinel15}.
 
In the case of rigid rotation, a global co-rotating frame exists such that
\begin{align}
    t' = t, \ \ \varphi'  =  \varphi - \Omega t, \ \ r' = r, \ \ z' = z, \label{eq:coordchan}
\end{align}
defines a coordinate transformation. In this rotating frame, the fluid's four-velocity takes the form 
$u'^{\mu} = \mathcal{F}^{-1}\delta^{\mu}_{\ t'}$, and the gauge potential transforms according to $A'^{\mu} = \Phi \delta_{\mu}^{\ t'} + \psi \delta_{\mu}^{\ \varphi '}$. 

Under the coordinate transformation \eqref{eq:coordchan}, the metric given by Eq. \eqref{eq:metric3} becomes
\begin{align}
 ds^2 = -F dt'^2 + 2K dt'd\varphi' + L d\varphi'^2 + e^{\mu} \big(dr'^2 + dz'^2\big),\label{eq:corotmetric}
\end{align}
where $F$ and $K$ are given by Eqs.~\eqref{eq:F} and~\eqref{eq:K} and $L = l$, with these quantities depending on the variables $r^\prime$ and $z^\prime$ only.  
For the remainder of this section, we will drop the primes.

Similarly, from relations \eqref{eq:rotE} and \eqref{eq:rotB}, we see that the electric and magnetic fields reduce to 
\begin{align}
    E_{\mu}  & = -\delta_{\mu}^{\ j} \mathcal{F}^{-1} \nabla_j \Phi, \label{eq:co-rotE}\\
    B_{\mu} & = - \frac{1}{2} \epsilon_{\mu \nu \alpha \beta} F^{\alpha \beta}  u^{\nu}  \\ & 
    =      - \delta_{\mu}^i \epsilon_{t i j \varphi}\frac{\mathcal{F}^{-1}}{D^2}\big(K \nabla^j \Phi + F \nabla^j \psi\big),  \label{eq:co-rotB}
\end{align}
respectively. 
Moreover, according to Eqs.~\eqref{eq:grhoel1} and \eqref{eq:grhomg2}, the electric and magnetic components of the electromagnetic energy density now read
\begin{align}
    & \rho_{el}  = \frac{1}{8 \pi F} \left(\nabla \Phi\right)^2, \label{eq:rhoel2} \\
    & \rho_{mg}  = \frac{1}{8\pi D^2 F}\big(K \nabla \Phi + F \nabla \psi\big)^2, \label{eq:rhomg2} 
\end{align}
respectively. 

Equation~\eqref{eq:relequi} assumes the form  \eqref{eq:relequi-rigid}, $ \left(\rho_m+p\right)\nabla_i\mathcal{F} +\rho_e\nabla_i\Phi  + \mathcal{F}\nabla_i p=0$,
indicating that the potentials $\mathcal{F}$ and $\Phi$ are functionally related to the pressure $p$. 

Finally, for rigid rotation, Eq.~\eqref{eq:generalsmarr} reduces to the much simpler form
\begin{align}
M - 2 \Omega J - \Phi Q = & -\int_{V_m}\left[\left(\rho_m + 3p\right)\mathcal{F} + \rho_e \Phi\right] u_{\mu} n^\mu dV \nonumber \\
 & + \Phi \int_{V_m} \rho_e u_{\mu}n^{\mu} dV,
\label{eq:smarr}
\end{align}
where the last four integral terms in Eq.~\eqref{eq:generalsmarr} have been eliminated due to the fact that $\Omega$ is a constant parameter.

\subsection{The QBH limit of rigidly rotating charged dust fluids }

\subsubsection{Rigidly rotating charged dust of Weyl type: linear relation}
\label{sec:rigidcd-linear}
 
We demonstrate here that a rigidly rotating, Weyl-type charged dust configuration, where the generalized electromagnetic potential $\Phi$ is linearly related to the redshift function $\cal F$, 
can attain the QBH limit while ensuring all relevant physical fields remain finite and well-defined.

Given that the fluid is pressureless and is in rigid rotation, equilibrium condition Eq.~\eqref{eq:relequi-rigid} takes the form 
\begin{equation}
    \rho_m d \mathcal{F} + \rho_e d \Phi  =0, \label{eq:equi-rigid-dust}
\end{equation}
with $d \mathcal{F} = \nabla_j \mathcal{F}dx^j$ and $d\Phi = \nabla_j \Phi dx^j$.  This relation indicates that the level surfaces of the redshift function $\mathcal{F}$ coincide with those of the electromagnetic potential $\Phi$, implying that $\mathcal{F}$ is functionally dependent on $\Phi$, i.e., $\mathcal{F} = \mathcal{F}(\Phi)$. As demonstrated in Ref.~\cite{basso2023}, this result represents the rigid rotation generalization of a theorem originally due to De and Raychaudhuri~\cite{deray68}, and further discussed in Ref.~\cite{Lemos:2009}. 

Moreover, from a suitable combination of the Einstein-Maxwell field equations, it is possible to obtain the following equation 
\begin{equation}
 \sqrt{\mathcal{F}'^2 - 1}\, \nabla_j \left(\sqrt{\mathcal{F}'^2 - 1}\, \nabla^j \Phi\right)    = 8 \pi  {\cal F} \mathcal{F}'  \rho_{mg} + \Xi_1,\label{eq:relpot12a}
\end{equation}
where  $\rho_{mg}$ is given in Eq.~\eqref{eq:rhomg2}, and $\Xi_1$ stands for 
\begin{equation}
\begin{aligned}
\Xi_1  =  & - \frac{\mathcal{F}'}{2 D^2{\cal F} F} \big(F \nabla K -  K \nabla F \big)^2 \\ & + \frac{1 }{D^2 F}\big(K \nabla F- F \nabla K \big) \cdot \big(K \nabla \Phi +F \nabla \psi \big).
\end{aligned} \label{eq:xi1}
\end{equation}

The last equation allows us to demonstrate the following result, originally presented in Ref.~\cite{basso2023}. Consider a rigidly rotating distribution of charged dust whose equipotential surfaces are closed, enclosing regions free of singularities or holes. If the right-hand side of Eq.~(\ref{eq:relpot12a}) vanishes under these conditions, then the redshift function $\mathcal{F}$ and the electrostatic potential $\Phi$  are linearly related as  
\begin{align}
    \mathcal{F} = - \epsilon \Phi + \gamma, \label{eq:lineara}
\end{align} 
where $\gamma$ is a constant of integration and $\epsilon = \pm 1$. As a consequence of this relation and the equilibrium equation \eqref{eq:equi-rigid-dust}, the charge and mass densities are proportional to each other,
\begin{equation}
\rho_e = \epsilon \rho_m.    \label{eq:lineara1}
\end{equation}

It is worth mentioning that this result does not assume any shape for the fluid and, therefore, can be used to generalize the QBH limit obtained in~\cite{Meinel15} for a disk composed of charged dust, as we shall now see.

Substituting Eqs.~\eqref{eq:lineara} and~\eqref{eq:lineara1} into Eq.~\eqref{eq:smarr}, and taking into account that $p=0$ in this case, it is straightforward to show that
\begin{align}
M - 2\Omega J - \Phi Q = \epsilon \gamma  Q - \Phi Q.  \label{eq:quasismarr}
\end{align}
Now, by taking the QBH limit $\Omega \to \omega_h$ and $\mathcal{F}^2 \to 0$, we find that $\Phi = \epsilon \gamma = \text{constant} \equiv \Phi_h$ everywhere in the interior region of the fluid. Hence, relation \eqref{eq:quasismarr} can be cast into the form 
 \begin{align} 
     M = 2 \omega_h J + \Phi_h Q.\label{eq:truesmarr} 
 \end{align} 

Equation \eqref{eq:truesmarr} is the well-known Smarr relation for an extremal Kerr--Newman black hole \cite{Smarr}. As demonstrated above, this relation also holds for extremal rotating QBHs composed of charged dust.
In \cite{Meinel15}, the authors showed that the QBH limit is attained if and only if the electromagnetic potential is constant on the disk. In contrast, we emphasize here that the constancy of this potential is a direct consequence of the QBH limit combined with Eq. \eqref{eq:lineara}.

It is worth noting that since the electromagnetic potential is constant throughout the inner region, its gradient vanishes inside the fluid, i.e., $\nabla_j \Phi = 0$. This implies that the rotating charged dust experiences a vanishing Lorentz force. Interestingly, in the absence of the Lorentz force, there are known cylindrically symmetric solutions~\cite{Som, Islam1977, Banerjee} characterized by $\rho_e = \epsilon \rho_m$ and by constant values of $\mathcal{F}$ and $\Phi$ that satisfy $\mathcal{F} = - \epsilon \Phi + \gamma$. However, in these solutions, the metric function $K$ and the magnetic potential $\psi$ are functionally related, which is not the case here. While a potential direction for future investigation would be to search the literature for rotating solutions that admit a QBH limit, the present work does not focus on any specific solution.

Another important point, which was not addressed in Ref.~\cite{Meinel15}, is whether the electromagnetic energy density remains finite in this limit. Thus, from Eq.~\eqref{eq:rhoel2}, the QBH limit implies that the ratio between $\nabla_j \Phi$ and $\mathcal{F}$ must remain finite for the electric energy density to remain finite as well. Since the relation between $\mathcal{F}$ and $\Phi$ is linear, it follows that $\Phi \to \Phi_h$ and $\nabla_j \Phi \to 0$ at the same rate as $\mathcal{F} \to 0$, i.e., $\nabla_i \Phi \sim \sqrt{\varepsilon}$. This gives
\begin{equation}
\rho_{el}  \sim \left(\frac{\nabla \Phi}{ 
\mathcal{F}}\right)^2 \sim \mathcal{O}(1), \label{eq:rhoelfinite}
\end{equation}
which ensures that the electric field and the corresponding energy density remain finite. 
The magnetic contribution is also regular. Indeed, using $K\to \sqrt{\varepsilon}$, $F\to\varepsilon$, $D^2 \to\varepsilon$, and $\nabla\psi\sim \mathcal{O}(1)$, we have
$ K\nabla\Phi+F\nabla\psi\to \varepsilon,$ and therefore Eq.~\eqref{eq:rhomg2} gives
\begin{equation}
    \rho_{mg}     \sim
    \Big(\frac{K\nabla\Phi+F\nabla\psi}{D\, \mathcal{F}}\Big)^2
    \sim \mathcal{O}(1). \label{eq:rhomgfinite}
\end{equation}
Thus, both the electric and magnetic fields, as well as their corresponding energy densities, remain finite in the QBH limit.

Finally, the same asymptotic behavior guarantees the consistency of Eq.~\eqref{eq:relpot12a}. In fact, ${\cal F}'{\cal F}\rho_{mg}\to \sqrt{\varepsilon}\to0$. Moreover, $F\nabla K-K\nabla F \to \varepsilon^{3/2}$ and $K\nabla\Phi+F\nabla\psi \to \varepsilon$, which imply $\Xi_1 \to \sqrt{\varepsilon}\to0$. Hence, the right-hand side of Eq.~\eqref{eq:relpot12a} vanishes in the QBH limit.

\subsubsection{Rigidly rotating charged dust of Weyl type: Islam ansatz}
\label{sec:rigidIslamdust}

We demonstrate here that for a rigidly rotating, Weyl-type charged dust satisfying the Islam ansatz (see below), the QBH limit can be reached only at the expense of arbitrarily large electromagnetic fields or fluid energy densities within the fluid distribution.

To facilitate the simplification of the field equations, motivated by the works of Islam \cite{Islam1978}, Bonnor \cite{Bonnor1980b}, and Raychaudhuri \cite{Raycha}, in Ref.~\cite{basso2023} we proposed the following ansatz connecting the metric and electromagnetic potentials,
\begin{align}
    & \nabla_j F = 2 \alpha \Phi \nabla_j \Phi, \ \ \  \nabla_j K = -2 \alpha \Phi \nabla_j \psi, \label{eq:ansatz1}
\end{align}
which we refer to as the Islam ansatz. One can readily verify that Eq.~(\ref{eq:ansatz1}) leads to the relation
\begin{equation}
F = \alpha \Phi^2 + \beta, \label{eq:FIslam}
\end{equation}
where $\alpha$ and $\beta$ are arbitrary constants. In the specific case where $\alpha =4$ and $\beta=0$, this expression reduces to the form proposed by Bonnor~\cite{Bonnor1980b}. Furthermore, when the Lorentz force vanishes, the ansatz given by Eq.~(\ref{eq:ansatz1}) becomes equivalent to the Bonnor-Raychaudhuri ansatz, which, in our notation, reads $u_\mu = \epsilon \sqrt{\alpha} A_\mu$. 

By adopting the ansatz \eqref{eq:ansatz1}, it is possible to show that the rigidly rotating charged dust satisfies the following relation~\cite{basso2023},
\begin{align}
    \big[\rho_m + 2\left(1 - \alpha\right) \rho_{el} + 2 \rho_{mg}\big]\mathcal{F} + \alpha \Phi \rho_e = 0. \label{eq:stateq}
\end{align}
This equation expresses the energy balance for configurations of rigidly rotating charged dust fluids that obey the Islam ansatz. The first term in Eq.~\eqref{eq:stateq}, which involves $\rho_m$, $\rho_{el}$, $\rho_{mg}$, and $\mathcal{F}$, corresponds to the gravitational energy density. Conversely, the second term, $\alpha \Phi \rho_e$, accounts for the electromagnetic binding energy of the system~\cite{Lemos:2009}.

Notice that when taking the QBH limit where $\Omega \to \omega_h$ and $F \to 0$, relation \eqref{eq:FIslam} implies that $\Phi\to \epsilon \sqrt{-\beta/\alpha}\equiv \Phi_h  = \text{constant}$ everywhere within the interior region of the fluid. As a consequence, the Lorentz force also vanishes in the fluid's interior, that is, $\nabla_j \Phi = 0$ throughout the fluid. If we require the fluid energy density $\rho_m$ and the electromagnetic energy densities $\rho_{el}$ and $\rho_{mg}$ to remain finite in this limit, Eq.~\eqref{eq:stateq} reduces to $\alpha \Phi_h \rho_e = 0$. This leads to three distinct scenarios: $\alpha = 0$, $\Phi_h = 0$, or $\rho_e = 0$.
In the case where $\alpha = 0$, relation \eqref{eq:FIslam} implies that $F=\beta$. The QBH limit $F\to 0$ then dictates that both the metric potential $F = \mathcal{F}^2$ and the electromagnetic potential $\Phi_h$ vanish trivially. 
The case $\Phi_h = 0$ means that the electromagnetic potential vanishes everywhere inside the fluid. Assuming it also vanishes at spatial infinity in the exterior region, consistently with the behavior of the electromagnetic potential in the Kerr--Newman geometry, it follows that the potential vanishes throughout the entire spacetime. This indicates the absence of a charged fluid from the outset.
Finally, the case $\rho_e = 0$ also trivially implies that no charged fluid is present.

Alternatively, combining Eqs.~\eqref{eq:equi-rigid-dust} and~\eqref{eq:FIslam} yields $\rho_e \mathcal{F} + \alpha \rho_m \Phi = 0$. From this relation, it follows that the QBH limit with bounded $\rho _{e}$ requires $\alpha = 0$, $\rho_m = 0$, or $\Phi_h = 0$. As in the previous paragraph, the cases $\alpha = 0$ and $\Phi_h = 0$ yield trivial solutions, whereas $\rho_m = 0$ implies the complete absence of matter throughout the spacetime.

Consequently, we arrive at the following conclusion: for a rigidly rotating charged dust of Weyl type satisfying the Islam ansatz \eqref{eq:ansatz1}, the extremal QBH limit cannot be attained under the assumption that the fluid and the electromagnetic energy density remain everywhere finite throughout the spacetime.

\subsection{The QBH limit of rigidly rotating charged pressure fluids }

\subsubsection{Rigidly rotating charged pressure fluids of Weyl type: Kloster--Das relation}
\label{sec:rigidKDas}

We show here that a rigidly rotating charged fluid with nonzero pressure where the potentials satisfy the Kloster--Das relation (see below) can attain the QBH limit while all relevant physical fields remain finite and well-defined. In this limit, the fluid becomes effectively pressureless.

Let us begin by considering the analysis of rigidly rotating charged pressure fluids. In such a case, the equilibrium equation (\ref{eq:relequi}) reduces to 
\begin{equation}
    \left(\rho_m + p\right)\, d \mathcal{F} + \mathcal{F}\, d p + \rho_e d \Phi = 0. \label{eq:equi-rigid-pressure}
\end{equation}
This relation indicates that the quantities $\mathcal{F}$, $\Phi$, and $p$ are not independent, but instead satisfy a functional dependence--for example, one may write $p = p(\mathcal{F}, \Phi)$.
In fact, the partial derivatives of $p$ obey 
\begin{align}
    \left( \frac{\partial p}{\partial \mathcal{F}} \right)_{\!\!\Phi} = -\frac{\rho_m + p}{\mathcal{F}},  \quad \left( \frac{\partial p}{\partial \Phi} \right)_{\!\!\mathcal{F}} = -\frac{\rho_e}{\mathcal{F}}.\label{eq:partial}
\end{align}
These functional relationships enable us to show that if any two of the level surfaces defined by constant values of $\mathcal{F}$, $\Phi$, and $p$ coincide, then the remaining one must necessarily coincide with them as well. This result was originally proposed by Guilfoyle \cite{Guilfoyle:1999yb} for the static case and generalized for the rigidly rotating case in Ref.~\cite{basso2023}. Kloster and Das~\cite{Kloster} were the first to highlight that, in the case of isometric motion, the pressure is an implicit  function of the potentials $\mathcal{F}$ and $ \Phi$ alone.

While studying stationary gravitational fields generated by dually charged fluids, Das and Kloster~\cite{Das1977} followed a slightly different route to find a particular class of solutions. Among other simplifying hypotheses, they assumed that the ratio of the charge density to the effective gravitational energy density of the fluid, $\rho_e / (\rho_m + 3p)$, remains constant throughout the fluid distribution. Since the term $(\rho_m + 3p)$ acts as the source for gravitational acceleration (as seen in the Tolman-Oppenheimer-Volkoff or Raychaudhuri equations), keeping this ratio constant ensures that the electrostatic repulsion scales with the gravitational attraction, thereby maintaining a static or equilibrium configuration. 

Inspired by Ref.~\cite{Das1977}, we consider here that the rotating charged fluid satisfies the following Kloster--Das relation
\begin{align}
    \frac{\rho_e}{\rho_m + 3p} = \epsilon \sqrt{\alpha}, \label{eq:Kloster}
\end{align}
where $\alpha$ is an arbitrary constant and $\epsilon = \pm 1$. In this case, it is easy to show, from Eq.~\eqref{eq:partial}, that 
\begin{align}
    p = \mathcal{F}^2 g\left(\epsilon \sqrt{\alpha}\, \Phi + \mathcal{F}\right)\label{eq:pressure}
\end{align}
with $g$ being an arbitrary differentiable function of $\epsilon \sqrt{\alpha} \Phi + \mathcal{F}$. 

Hence, we obtain the following result. If a rigidly rotating charged fluid with pressure satisfies Eq.~\eqref{eq:Kloster}, then $p = w \mathcal{F}^2$ if, and only if, $\mathcal{F}$ is of the form
\begin{equation}
    \mathcal{F} = \sqrt{\alpha} (- \epsilon\Phi + \gamma), \label{eq:KDlinear}
\end{equation}
where $w$ and $\gamma$ are arbitrary constants. This fact follows immediately from Eq.~\eqref{eq:pressure} for $w = g(\sqrt{\alpha}\gamma)$. Interestingly, this result constitutes a generalization to the rigidly rotating case of a theorem originally established by Guilfoyle~\cite{Guilfoyle:1999yb} for the static case. It is worth mentioning that this fact was not noticed in Ref.~\cite{basso2023}. Moreover, Eq.~\eqref{eq:smarr} reduces once more to Eq.~\eqref{eq:quasismarr}.

By taking the QBH limit, where $\Omega \to \omega_h$ and $\mathcal{F}^2 \to 0$, it follows that $p = w\mathcal{F}^2 \to 0$ and $\Phi = \epsilon \gamma = \text{constant} \equiv \Phi_h$ throughout the interior region. Consequently, the Smarr relation~\eqref{eq:truesmarr} for an extremal Kerr--Newman geometry is recovered. Thus, Eq.~\eqref{eq:truesmarr} holds for the QBH limit of a rigidly rotating, charged fluid that satisfies the Kloster--Das relation. Notably, in this case, the charged fluid becomes pressureless in the QBH limit.

Since $\Phi\to\Phi_h$ throughout the interior region, its gradient vanishes in the QBH limit and, consequently, so does the Lorentz force in the comoving frame. Moreover, the regularity analysis is
analogous to that presented in Sec.~\ref{sec:rigidcd-linear}. Indeed, the linear relation between $\mathcal{F}$ and $\Phi$, cf. Eq.~\eqref{eq:KDlinear}, implies that $\nabla_i\Phi \to \sqrt{\varepsilon}$. Hence, from Eq.~\eqref{eq:rhoel2}, one has Eq.~\eqref{eq:rhoelfinite}. Similarly, using the limiting behavior of $K \to \sqrt{\varepsilon}$,
$F \to \varepsilon$, and $D^2 \to \varepsilon$, it follows that Eq.~\eqref{eq:rhomg2} approaches the same limiting value for $\rho_{mg}$ as in the previous subsection, as shown in Eq.~\eqref{eq:rhomgfinite}. Therefore, both the electric and magnetic fields, as well as their corresponding energy densities, remain finite in the QBH limit.

\subsubsection{Rigidly rotating charged pressure fluids of Weyl type: Guilfoyle relation}
\label{sec:rigidguilfoyle}

We show here that the QBH limit can also be consistently achieved for rigidly rotating charged fluids with pressure, provided that the potentials satisfy the Guilfoyle relation, 
 \begin{align}
     {\cal F}^2 = \alpha\left( -\epsilon \Phi + \gamma\right)^2 + \beta,\label{eq:guilfoyle}
\end{align}
where $\alpha$, $\gamma$, and $\beta$ are arbitrary constant parameters. 

A few interesting special cases emerge from the Guilfoyle relation \eqref{eq:guilfoyle}. First, setting $\alpha = 1$ and $\beta = 0$ yields the linear relationship between $\mathcal{F}$ and $\Phi $ that characterizes the rigidly rotating charged dust studied in Sec.~\ref{sec:rigidcd-linear}, where the QBH limit is well-behaved. Second, for $\alpha = 1$ with nonzero $\gamma $ and $\beta $, the original relation obtained by Weyl \cite{Weyl} is recovered. As shown by Guilfoyle \cite{Guilfoyle:1999yb} for non-rotating charged fluids, if there is a closed boundary surface where the pressure goes to zero, then this relation implies that the pressure vanishes everywhere. While it is not straightforward to verify whether this result also holds for rotating charged fluids, this case matches the QBH limit features of the general case where $\alpha \neq 1$ and $\beta \neq 0$ that we investigate below. Third, for $\alpha \neq 1$ and $\beta = 0$, one obtains a linear relation between $\mathcal{F}$ and $\Phi $, namely $\mathcal{F} = \sqrt{\alpha}\left(-\epsilon\Phi + \gamma\right)$. The existence of the QBH limit for spherically symmetric, static, charged fluids satisfying this relation was demonstrated in Ref.~\cite{Lemos:2010te}. 

Next, we examine the QBH limit for the general Guilfoyle relation with $\alpha \neq 1$, considering both $\beta \neq 0$ and $\beta = 0$, and identify the corresponding consistency conditions.

Focusing on the field equations for a rigidly rotating charged fluid with pressure, we find (see Ref.~\cite{basso2023} for more details)
\begin{align}
    & \nabla_j \Big[\frac{1}{\mathcal{F}} \nabla^j \Big(F - \alpha( -\epsilon \Phi + \gamma)^2  - \beta\Big) \Big]  = \Xi_2 \,+ \label{eq:relpot17} \\ 
    & 8 \pi \Big[ \big( \rho_m  + 3 p  + 2(1 - \alpha) \rho_{el} + 2 \rho_{mg}\big)\mathcal{F}
    +  \epsilon \alpha \left(\epsilon\Phi - \gamma\right)  \rho_e  \Big]. \nonumber
\end{align}
For notational convenience, the symbol $\Xi_2$ in Eq.~\eqref{eq:relpot17} is used to represent
\begin{equation}
\begin{aligned} 
\Xi_2 = &  \frac{\mathcal{F}}{F^2D^2}\Big[2\alpha\Phi\big(K \nabla F - F \nabla K\big)\!\cdot\!\big(K \nabla \Phi + F \nabla \psi\big)\\ &  -  \big(K \nabla F - F \nabla K\big)^2\Big].\label{eq:Xi}
\end{aligned}
\end{equation}

The relation \eqref{eq:relpot17} allowed us to generalize a theorem from~\cite{Lemos:2009} which can be stated as follows. Consider a rigidly rotating distribution of charged fluid with pressure whose equipotential surfaces are closed, enclosing regions free of singularities or holes. Then, if the right-hand side of Eq.~(\ref{eq:relpot17}) vanishes under these conditions, the redshift function $\mathcal{F}$ and the electrostatic potential $\Phi$ are related by the Guilfoyle relation \eqref{eq:guilfoyle}. On the other hand, if the constraint $F - \alpha\left( -\epsilon \Phi + \gamma\right)^2 - \beta = 0$ holds, then it follows that the right-hand side of Eq.~(\ref{eq:relpot17}) also vanishes.  

Because the Guilfoyle relation~\eqref{eq:guilfoyle} is assumed to hold {\it a priori}, the left-hand side of Eq.~\eqref{eq:relpot17} vanishes identically, requiring the right-hand side to vanish everywhere as well.  With this well-established situation, we need to analyze the QBH limit on the right-hand side of such an equation. Assuming that the densities of matter and electromagnetic energy 
remain finite, the term proportional to $\mathcal{F}$ in Eq.~\eqref{eq:relpot17} vanishes as
$\mathcal{O}(\sqrt{\varepsilon})$. Therefore, in the QBH limit,
Eq.~\eqref{eq:relpot17} gives
\begin{equation}
    \Xi_2 + 8 \pi\epsilon\alpha(\epsilon\Phi_h-\gamma)\rho_e
    +\mathcal{O}(\sqrt{\varepsilon}) = 0. \label{eq:relpot17limit}
\end{equation}
Taking into account that the Guilfoyle relation~\eqref{eq:guilfoyle} in the QBH limit implies that
\begin{equation}
    \epsilon\Phi_h-\gamma
    =  \epsilon\sqrt{-\beta/\alpha},
\end{equation}
which is finite and well-defined for $\alpha\neq0$ and $-\beta/\alpha>0$, Eq.~\eqref{eq:relpot17limit} leads to $ \Xi_{2} \to - 8\pi
\sqrt{-\alpha \beta}\, \rho_{e}.$ Hence, the parameters $\alpha$ and $\beta$ fix the finite contribution required
to balance $\Xi_2$ in the QBH limit, ensuring the consistency of
Eq.~\eqref{eq:relpot17}. In other words, 
\begin{equation}
   \frac{\Xi_2^2}{\rho_e^2} \to -64 \pi^2 \alpha \beta = \text{constant}
\end{equation}
should be understood as a consistency condition in the QBH limit.

The case $\beta = 0$ is also compatible with this limiting relation, provided that $\Xi_2 \to 0$ in the QBH limit, yielding $\Phi \to \Phi_h = \epsilon\gamma$. This is expected because setting $\beta = 0$ in the general Guilfoyle relation \eqref{eq:guilfoyle} recovers the linear relation between ${\cal F}$ and $\Phi$ used in the previous subsection.

Now, in order to obtain the Smarr relation, let us follow Ref.~\cite{Lemos2008} and notice that
\begin{align}
    -\int_{V_m}\!\!\!\left(\rho_m + 3p\right)\mathcal{F} u_{\mu} n^{\mu}d V \le - \mathcal{F}_{\!B}\! \int_{V_m}\!\!\!\left (\rho_m + 3p\right)u_{\mu} n^{\mu}d V, \label{eq:masslimit}
\end{align}
where $\mathcal{F}_B$ is the redshift factor at the boundary of the fluid. Since $u_{\mu}n^{\mu} <0$ for future-directed timelike vectors, the left-hand side of Eq.~\eqref{eq:masslimit} is nonnegative provided that $\rho_m + 3p \ge 0$. 
In the interior region, when approaching the boundary of the fluid distribution from the inside, the QBH limit is characterized by $\mathcal{F} \le \mathcal{F}_B \to 0$. For a finite $\rho_m + 3p \ge 0$, the left-hand side of Eq.~\eqref{eq:masslimit} vanishes in this limit. Consequently, Eq.~\eqref{eq:smarr} reduces to the Smarr relation~\eqref{eq:truesmarr} for an extremal Kerr--Newman geometry, provided that the electromagnetic potential becomes constant everywhere inside the fluid and can be pulled out of the integral. Thus, Eq.~\eqref{eq:truesmarr} is established in the QBH limit of a rigidly rotating, charged fluid with pressure that satisfies the Guilfoyle relation.

Since in the QBH limit the electromagnetic potential $\Phi \to \Phi_h$ is constant throughout the entire inner region, its gradient must vanish inside the fluid. As a result, the Lorentz force acting on the rigidly rotating charged fluid is zero. Moreover, for the same reasons discussed in Sec.~\ref{sec:rigidcd-linear}, the electromagnetic energy and the electromagnetic field remain finite in the QBH limit.

\subsubsection{Rigidly rotating charged pressure fluids of Weyl type: Islam ansatz}

We now examine the implications of the Islam ansatz, as given in Eqs.~\eqref{eq:ansatz1}, for a rotating charged fluid of Weyl type with pressure. As already mentioned above, it is straightforward to verify that $\Xi_2 = 0$, $F = \alpha \Phi^2 + \beta$ [see Eq.~\eqref{eq:FIslam}]. With this, Eq.~(\ref{eq:relpot17}) implies that
\begin{equation}
\left[\rho_m + 3 p + 2\big(1 - \alpha\big) \rho_{el} + 2 \rho_{mg}\right]\mathcal{F} + \alpha \Phi \rho_e = 0.  \label{eq:statef}
\end{equation}

Similarly to the case of zero pressure analyzed in Sec.~\ref{sec:rigidIslamdust}, in this case the extremal QBH limit cannot be reached if one requires the energy density, pressure, and electromagnetic energy to remain finite. To prove this statement, one may follow the same steps as in Sec.~\ref{sec:rigidIslamdust}, and we do not present it here. The nonexistence of the regular QBH limit for the rotating charged pressure fluids that satisfy the Islam ansatz is consistent with the static case analyzed in Ref.~\cite{Lemos:2010te}.
In that case, the authors examined the solutions obtained by Guilfoyle~\cite{Guilfoyle:1999yb} for static charged pressure fluids, and found that the extremal QBH limit is achieved only for the class of solutions with $\beta = 0$ and $\gamma \neq 0$ in the Guilfoyle relation~\eqref{eq:guilfoyle}.

\section{Quasiblack-hole limit of differentially rotating charged fluids}
\label{sec:differential}

\subsection{General properties}

We discuss here the QBH limit of differentially rotating charged fluid configurations.

As shown in Sec.~\ref{sec:qbhlimit}, the fluid's angular velocity $\Omega$ must approach a constant value $\omega_h$ throughout the inner region in this limit, even if the rotation is not rigid initially. Although the combination in Eq.~\eqref{eq:chi} ceases to define a Killing vector field under differential rotation, the relation $\chi_{\mu}\chi^{\mu} =- \mathcal{F}^2$ still holds and, as the QBH limit is approached, $\chi^{\mu}$ becomes a null Killing vector field everywhere within the inner region.

\subsection{Differentially rotating charged fluids with vanishing Lorentz force}
\label{sec:VB}

We show here that the QBH limit can be achieved for differentially rotating charged fluids for vanishing Lorentz force with the electromagnetic field remaining finite.

When analyzing the properties of differentially rotating charged fluid configurations within the Einstein-Maxwell framework, early studies relied on simplifying assumptions to bypass the complexity of the full set of equations. In particular, these initial works often assumed zero-pressure fluids with a vanishing Lorentz force. The search for exact solutions describing such differentially rotating charged dust fluids was pioneered by Islam \cite{Islam1978, Islam1979, Islam1983}, as well as by Van den Bergh, Wils, and Islam \cite{Bergh, Wils, Islam1984}.

In the special case where the Lorentz force vanishes, the condition $\nabla_j \Phi - \psi \nabla_j \Omega = 0$ holds, implying that the electromagnetic potential $\Phi$ depends solely on the angular velocity, i.e., $\Phi = \Phi(\Omega)$ with $d\Phi/d\Omega = \psi$. Since our focus here is not on constructing explicit solutions, we proceed to analyze the QBH limit of such configurations. Let us note here that, for systems with vanishing Lorentz force, this analysis can be performed without assuming the  Weyl-type ansatz, i.e., the assumption $\mathcal{F} = \mathcal{F}(\Phi)$ is not necessary here.

In the QBH limit where $\Omega \to \omega_h$ (with $\omega_h$ being a constant parameter), the condition of a vanishing Lorentz force implies that $\Phi$ is also constant throughout the interior of the fluid, i.e., $\Phi \to \Phi_h = \text{const}$. Furthermore, the conditions imposed on the sequence of functions $\Omega_n$ and $\omega_n$ as $\Omega \to \omega_h$ ensure that the last two terms on the right-hand side of Eq.~\eqref{eq:generalsmarr} cancel out. These conditions also guarantee that $\nabla_j \Omega \to 0$ throughout the interior region, thereby eliminating the terms involving the gradient of $\Omega$ in the same equation. Following the reasoning in Sec.~\ref{sec:rigidguilfoyle}, the left-hand side of Eq.~\eqref{eq:masslimit} vanishes as well. Consequently, once the electromagnetic potential becomes constant everywhere inside the fluid and can be pulled outside the integral, Eq.~\eqref{eq:generalsmarr} reduces to the Smarr relation~\eqref{eq:truesmarr} for an extremal Kerr--Newman geometry. Unlike the rigidly rotating configurations discussed in the previous section, here the mass relation~\eqref{eq:truesmarr} is obtained in the QBH limit for a differentially rotating, charged pressure fluid with a vanishing Lorentz force.

In the absence of the Lorentz force, the equilibrium equation~\eqref{eq:relequi} reduces to
$ (\rho_m + p) \nabla_i \mathcal{F} + (\rho_m + p) \mathcal{F}^{-1} K \nabla_i \Omega + \mathcal{F}\,\nabla_i p = 0$. In the QBH limit, $\nabla_i \mathcal{F} \to 0$, $\nabla_i \Omega \to 0$, and $\mathcal{F} \to 0$, while the term $\mathcal{F}^{-1} K$ remains finite. Therefore, assuming that the matter variables $\rho_m$, $p$, $\rho_e$, and $\nabla_i p$ remain finite, all terms in the equilibrium equation vanish consistently in this limit. Consequently, no additional asymptotic condition on the pressure gradient is required, allowing the pressure within the fluid to remain non-zero in the QBH limit.
 
Finally, let us consider the electromagnetic energy densities.
As discussed in Sec.~\ref{sec:qbhlimit}, their regularity is controlled by the combinations entering Eqs.~\eqref{eq:grhoel1} and
\eqref{eq:grhomg2}. In the present case, the vanishing Lorentz
force condition holds identically throughout the fluid region, and then Eq.~\eqref{eq:grhoel1} gives $\rho_{el}=0$. Moreover, Eq.~\eqref{eq:grhomg2} reduces to $ \rho_{mg} = \frac{F}{8\pi D^2}\big(\nabla\psi\big)^2$, which, in the QBH limit tends to 
\begin{equation}
    \rho_{mg} \to  \frac{\varepsilon}{8\pi \varepsilon }\big(\nabla\psi\big)^2 \to {\rm finite}.
\end{equation}
Hence, the electromagnetic field and the respective energy density are regular in the QBH limit.

\subsection{Differentially rotating charged pressure fluids of Weyl type}
\label{sec:VC}
We demonstrate here that the QBH limit can also be consistently attained for differentially rotating charged fluids with pressure satisfying the Weyl ansatz, with all relevant electromagnetic fields remaining finite provided $\delta^2$ scales linearly with $\varepsilon$.

Let us now focus on the general features of differentially rotating charged fluids with
nonvanishing pressure satisfying the Weyl ansatz, i.e., such that the redshift factor ${\cal F}$  is functionally related to the generalized electromagnetic potential $\Phi$, ${\cal F} = {\cal F}(\Phi)$. This functional relation has an important
consequence in the QBH limit. The two contributions to the Lorentz-force term $\nabla_j\Phi-\psi\nabla_j\Omega$ vanish separately. Indeed, as discussed in Sec.~\ref{sec:qbhlimit}, in this limit
one has ${\cal F}\to \sqrt{\varepsilon}$ and $\nabla_i{\cal F} \to\sqrt{\varepsilon}$.  Moreover,  the Weyl ansatz gives $\nabla_i{\cal F}= {\cal F}'\nabla_i \Phi$ and then, for a well-behaved ${\cal F}' \neq 0$, one has that $\nabla_i{\Phi }$ goes to zero at the same rate as $\nabla_i{\cal F}$, i.e., $\nabla_i\Phi \to\sqrt{\varepsilon}$, assuring that the first term in the Lorentz force goes to zero. The assumptions imposed on the QBH behavior of
the angular velocity and its gradient likewise yield
$\psi\nabla_i\Omega\to0$, as summarized in
Table~\ref{tab:qbh-fluid-limit}. Consequently, $ \nabla_j\Phi-\psi\nabla_j\Omega\to0,$
throughout the fluid interior in the QBH limit.

In the setting for Weyl-type charged rotating fluids, the equilibrium condition~(\ref{eq:relequi}), expressed using total differentials, takes the form
\begin{align}
      &   \left[  \big(\rho_m + p\big) \mathcal{F}' + \rho_e\right] d \Phi + \mathcal{F} d p \nonumber  \\
     &   \qquad  \qquad   \qquad +  \left[\big (\rho_m + p \big)\mathcal{F}^{-1}K -  \rho_e \psi \right] d \Omega = 0, \label{eq:relequi1}
\end{align}
where $\mathcal{F}' = d \mathcal{F}/d\Phi$.

In analogy to the static case and to simplify the analysis, we assume the fluid quantities satisfy the relation  $(\rho_m + p)\mathcal{F}' + \rho_e = 0$. As shown in Ref.~\cite{basso2023}, under such an assumption, the surfaces of constant angular velocity necessarily coincide with those of constant pressure, i.e., the pressure $p$ is a function of the angular velocity $\Omega$ alone. Another immediate consequence is that the ratio $\rho_e / (\rho_m + p) = -\mathcal{F}'$ depends only on the electrostatic potential $\Phi$. This functional dependence enables a simplification of the system, allowing for direct relations among the fluid quantities alone. 
A particularly interesting case results when we make a further simplification by assuming the linear Weyl--Guilfoyle ansatz $\mathcal{F} = \sqrt{\alpha}(-\epsilon\Phi + \gamma)$, where $\alpha$ and $\gamma$ are constants. Notice that this is the same ansatz as in Eq.~\eqref{eq:guilfoyle} in the case $\beta=0$.
Under this assumption, one obtains the simple relation $\rho_e = \epsilon\sqrt{\alpha}(\rho_m + p)$, and the QBH limit may be obtained as follows.

In the QBH limit one has the limits $\mathcal{F} \to 0$, $\Omega\to \omega_h$, and $\Phi= \Phi_h$.
The limit $\mathcal{F} \to 0$ implies that $\Phi = \epsilon \gamma \equiv \Phi_h={\rm constant}$, and the conditions imposed on the sequence of the functions $\Phi_n$ in the limit $\Phi \to \Phi_h$ imply that $\nabla_j \Phi\to 0$.
These results ensure that the second and the third integral terms on the right-hand side of Eq.~\eqref{eq:generalsmarr} cancel out, thereby eliminating the terms involving $\Phi$ in that equation.  Similarly, the conditions imposed on the sequence of functions $\Omega_n$ and $\omega_n$ in the limit $\Omega \to \omega_h$ imply that $\nabla_j \Omega\to 0$. These results,  together with the condition $\nabla_j \Omega \to 0$ in the fluid region,  ensure that the last four integral terms on the right-hand side of Eq.~\eqref{eq:generalsmarr} also cancel out, thereby eliminating the terms involving the gradient of $\Omega$ and $\Phi$ in that equation. By the same reasoning given in Sec.~\ref{sec:rigidguilfoyle}, the first integral term on the right-hand side of Eq.~\eqref{eq:generalsmarr} also vanishes. Therefore, Eq.~\eqref{eq:generalsmarr} reduces to the Smarr relation~\eqref{eq:truesmarr} for an extremal Kerr--Newman geometry, but now, such a mass formula is obtained for the QBH limit of differentially rotating charged pressure fluids of Weyl type that satisfies $\mathcal{F} = \sqrt{\alpha}(-\epsilon\Phi + \gamma)$.

Finally, let us consider the regularity of the electromagnetic energy densities in the QBH limit. As discussed in Sec.~\ref{sec:qbhlimit}, ensuring the finiteness of Eq.~\eqref{eq:grhoel1} requires $\nabla_i\Phi - \psi\nabla_i\Omega \to \sqrt{\varepsilon}$. For the present Weyl-type configurations, the linear relation between $\mathcal{F}$ and $\Phi$ gives $\nabla_i\Phi \to \sqrt{\varepsilon}$, whereas $\nabla_i\Omega \to \sqrt{\varepsilon} + \delta$. Since we assume $\psi \sim \mathcal{O}(1)$, this regularity condition is satisfied provided $\delta \sim \mathcal{O}\left(\sqrt{\varepsilon}\right)$. Under this condition, Eq.~\eqref{eq:grhoel1} implies that $\rho_{el}$ is finite. Moreover, applying the regularity analysis of Sec.~\ref{sec:qbhlimit} to Eq.~\eqref{eq:grrhomg22} shows that $\rho_{mg}$ is also finite. Consequently, both the electromagnetic fields and their corresponding energy densities remain finite in the QBH limit.

\subsection{Differentially rotating charged fluids: general case}

The assumptions introduced in Sec.~\ref{sec:qbhlimit} allow us to establish the constancy of the electromagnetic potential in the QBH limit without imposing a Weyl-type relation or an {\it ad hoc}  vanishing Lorentz force. To this end, consider a connected distribution of charged fluid, possibly undergoing differential rotation, that satisfies the regularity and convergence conditions specified therein.

From Eq.~\eqref{eq:grhoel1}, the finiteness of the electric energy density requires the Lorentz force to vanish, thereby implying Eq.~\eqref{eq:lorentzforce}. Furthermore, the convergence assumptions of Sec.~\ref{sec:qbhlimit} dictate that $\nabla_i\Omega\to0$ while $\psi$ remains bounded. Consequently,
\begin{align}
  \nabla_i\Phi & = \left(\nabla_i\Phi-\psi\nabla_i\Omega\right)
  +\psi\nabla_i\Omega \to 0. \label{eq:nablaPhi-limit}
\end{align}
Hence, under the convergence assumptions of Sec.~\ref{sec:qbhlimit}, the asymptotic behavior of $\nabla\Phi$ implies that the limiting potential has a vanishing spatial gradient and is therefore constant throughout the connected fluid interior, $\Phi\to\Phi_h=\text{constant}$. Thus, within the regular class of fluid distributions and geometries considered here, the constancy of the limiting electromagnetic potential $\Phi$ follows directly from the regularity of the electromagnetic field and the vanishing of differential rotation in the specified sense. Notably, neither a functional relation between $\mathcal{F}$ and $\Phi$ nor an additional restriction on the charge-to-mass density ratio is required. 

Equation \eqref{eq:nablaPhi-limit} implies that for rigidly rotating charged fluids, the finiteness of the electric energy density directly constrains the gradient of the electromagnetic potential, yielding $\nabla_i\Phi\to0$. In contrast, for differentially rotating fluids, the corresponding regularity condition constrains the combination $\nabla_i\Phi-\psi\nabla_i\Omega$. Consequently, establishing $\nabla_i\Phi\to0$ requires controlling the term $\psi\nabla_i\Omega$. This is guaranteed by the assumptions in Sec.~\ref{sec:qbhlimit}, which ensure that $\nabla_i\Omega\to0$ and that $\psi$ remains bounded. Together with the assumed convergence of the potentials and their derivatives, these conditions establish the constancy of the limiting generalized electromagnetic potential throughout the fluid interior, without requiring a Weyl-type relation. The present work extends the electrostatic analysis of Lemos and  Zaslavskii~\cite{Lemos2008}, who demonstrated that the regularity of the electromagnetic energy density enforces the uniformity of the electrostatic potential on the quasihorizon and derived its interior limiting behavior. Although their subsequent treatment of rotating QBHs incorporated electromagnetic contributions via the quasihorizon potential~\cite{Lemos:2009wj}, the present analysis generalizes these arguments to stationary axisymmetric configurations, including differentially rotating fluids, by identifying the regularity condition on the generalized potential and establishing its constancy across the fluid interior.

The extremal Smarr relation then follows by taking into account the limiting behavior $\Phi\to\Phi_h$ and $\Omega\to\omega_h$, employing the condition in Eq.~\eqref{eq:masslimit}, and applying the same reasoning used for rigidly rotating Guilfoyle fluids in Sec.~\ref{sec:rigidguilfoyle}. Under the convergence and integrability assumptions inherent to that approach, the redshift-weighted matter contribution vanishes. The remaining terms on the right-hand side of Eq.~\eqref{eq:generalsmarr} either cancel out or vanish in the QBH limit, yielding Eq.~\eqref{eq:truesmarr}. Consequently, the extremal Smarr relation is recovered for the general class of charged fluids considered here, without requiring a Weyl-type relation or a vanishing Lorentz force \textit{a priori}.

\section{Final remarks}
\label{sec:conclusion}

\begin{table*}[t]
\caption{
Summary of the rotating charged-fluid configurations analyzed in this
work and their behavior in the extremal quasiblack-hole limit.
Here, a regular limit requires the matter and electromagnetic energy
densities to remain finite throughout the matter distribution region.
}
\label{tab:qbh-model-summary}

\centering
\scriptsize
\renewcommand{\arraystretch}{1.25}
\setlength{\tabcolsep}{2.5pt}

\begin{tabularx}{\textwidth}{
  @{}
  L{1.15cm}
  L{1.05cm}
  Y
  L{1.15cm}
  Y
  L{1.45cm}
  L{0.90cm}
  @{}
}
\toprule
Rotation
&
Matter
&
Relation between metric and EM potentials
&
\shortstack[l]{Regular\\QBH}
&
Regularity condition
&
\shortstack[l]{Pressure in\\the limit}
&
Smarr
\\
\midrule

Rigid
&
Dust
&
Linear Weyl relation:
$\mathcal{F}=-\epsilon\Phi+\gamma$
&
Yes
&
Automatically satisfied by the linear Weyl relation
&
$p=0$
&
Yes
\\

Rigid
&
Dust
&
Islam ansatz:
$\nabla_i F=2\alpha\Phi\,\nabla_i\Phi$ and
$\nabla_iK=-2\alpha\Phi\,\nabla_i\psi$
&
No
&
Finite matter and electromagnetic energy densities require
a trivial configuration or obstruct the QBH limit
&
$p=0$
&
---
\\

Rigid
&
Pressure
&
Kloster--Das relation:
$\rho_e/(\rho_m+3p)=\epsilon\sqrt{\alpha}$,
with $\mathcal{F}=\sqrt{\alpha}(-\epsilon\Phi+\gamma)$
&
Yes
&
Automatically satisfied by the linear relation
&
$p\rightarrow0$
&
Yes
\\

Rigid
&
Pressure
&
Guilfoyle relation:
$\mathcal{F}^{2}
=\alpha(-\epsilon\Phi+\gamma)^{2}+\beta$
&
Yes
&
$\rho_m+3p\geq0$ and $\Xi_2/\rho_e \to \text{constant}$
&
May remain nonzero
&
Yes
\\

Rigid
&
Pressure
&
Islam ansatz
&
No
&
Finite matter and electromagnetic energy densities require
a trivial configuration or obstruct the QBH limit
&
---
&
---
\\

Diff.
&
Pressure
&
Vanishing Lorentz force:
$\nabla_i\Phi-\psi\nabla_i\Omega=0$
&
Yes
&
Automatically satisfied by the vanishing Lorentz force condition
&
May remain nonzero
&
Yes
\\

Diff.
&
Pressure
&
$(\rho_m+p)\mathcal{F}'+\rho_e=0$ with
$\mathcal{F}=\sqrt{\alpha}(-\epsilon\Phi+\gamma)$
&
Yes
&
$\delta \sim \mathcal{O}(\sqrt{\varepsilon})$
&
May remain nonzero
&
Yes
\\

\bottomrule
\end{tabularx}
\end{table*}

In this work, we have investigated the QBH limit of rotating charged fluids. By considering both rigidly and differentially rotating configurations, and by analyzing cases with and without pressure, we have shown that, under suitable conditions, such systems can attain the QBH limit. In this limit, the interior region is characterized by an infinite redshift, the quasi-horizon becomes a null hypersurface, and the exterior geometry approaches that of an extremal Kerr--Newman black hole. A summary of the main geometrical and physical properties of the particular cases in the QBH limit is presented in Tab. \ref{tab:qbh-model-summary}.

For rigidly rotating charged dust fluids satisfying a linear relation between the metric and electromagnetic potentials, the QBH limit implies a constant electromagnetic potential, vanishing Lorentz force, and finite electromagnetic energy. The Smarr formula for an extremal Kerr--Newman black hole is recovered. However, when the system obeys the so-called Islam ansatz, the QBH limit cannot be reached without diverging electromagnetic energy, highlighting the restrictive nature of this condition.

For rigidly rotating charged fluids with pressure, we have explored two relevant cases. Fluids satisfying the Kloster--Das relation become pressureless in the QBH limit, while those satisfying the Guilfoyle relation retain nonzero pressure. In both cases, the electromagnetic potential becomes constant, the Lorentz force vanishes, and the Smarr relation is again satisfied, provided suitable conditions are met to ensure the finiteness of the electromagnetic energy.

We have also extended our analysis to differentially rotating charged fluids. In cases where the Lorentz force vanishes, the QBH limit leads to a constant electromagnetic potential and allows for the recovery of the Smarr formula. Finally, for differentially rotating charged pressure fluids of Weyl type, the QBH limit is attainable under a linear ansatz between the potentials, with all physical conditions analogous to the rigidly rotating case.

In conclusion, our results demonstrate that rotating Weyl-type systems admit an extremal QBH limit in much the same way as their static counterparts. Moreover, these results reinforce and generalize previous findings on QBHs and support the idea that extremal Kerr--Newman geometry represents the universal exterior configuration approached by rotating charged fluids in the QBH limit. This observation mirrors a similar discussion in the context of uncharged fluids, namely, that the extremal Kerr black hole is the only candidate for the black-hole limit of rotating fluid bodies in equilibrium~\cite{Meinel02, Meinel04, Meinel06}. In the present charged case, our findings suggest that the extremal Kerr--Newman black hole appears to be the unique candidate for the black-hole limit of rotating charged fluids in equilibrium, provided the configurations satisfy the regularity conditions assumed throughout this work.

Our analysis here has been restricted to regular QBH limits in which the relevant physical quantities remain finite and the interior and exterior geometries are smoothly matched, without any surface layer at the quasihorizon. This assumption is consistent with the general properties of QBHs discussed in Refs.~\cite{Lemos:2009wj, Lemos2010}, where the requirement of finite surface stresses is associated with the extremal limit. A more general possibility arises if this assumption is relaxed. Indeed, Ref.~\cite{Lemos:2009wj} conjectured that, by allowing divergent surface stresses at the quasihorizon, stationary QBH configurations could approach a generic, not necessarily extremal, Kerr--Newman exterior. Such configurations lie outside the regular class investigated here. It would therefore be interesting to investigate whether rotating charged fluids can realize such nonextremal QBH limits in the presence of a surface layer.

As a natural continuation of this work, it would be important to search for explicit solutions realizing the QBH limit within families of rotating charged fluid solutions already known in the literature, as well as to construct new configurations that exhibit this behavior, beyond the charged dust disk solution presented in~\cite{Meinel13, Meinel15}. In that work, the QBH limit was shown to be attained if, and only if, the electromagnetic potential is constant on the disk. In contrast, here we find that the constancy of the electromagnetic potential follows naturally from the QBH limit together with Eq.~\eqref{eq:lineara}. This indicates that this property may be a more general feature of rotating charged QBH configurations and motivates further investigations in broader classes of solutions.

\begin{acknowledgments}

M. L. W. B. is funded by the Funda\c c\~ao de Amparo \`a Pesquisa do Estado de S\~ao Paulo (FAPESP), Brazil, Grant No.~2022/09496-8 and Grant No.~2025/07325-0. V. T. Z. is funded in part by Conselho Nacional de Desenvolvimento Cient\'ifico
e Tecnol\'ogico (CNPq), Brazil, Grants No.~311726/2022-4, and No.~309247/2026-8. We thank J.~P.~S.~Lemos for stimulating discussions.

\end{acknowledgments}

\appendix 

\section{Main steps for the derivation of Eq.~(38)}
\label{sec:appendix}

In this appendix, we describe the main steps leading to Eq.~\eqref{eq:generalsmarr}. Equation~\eqref{eq:smarr} then follows immediately as a particular case. The starting point is Eq.~\eqref{eq:mJ}, where the integral terms on the right-hand side of such an equation must be rearranged appropriately.

Let us begin by noting that the first term on the right-hand side of Eq.~\eqref{eq:mJ} can be written as
\begin{equation} \label{eq:mJap}
\begin{aligned}
     & 2 \int_{\Sigma_t}\Big(T_{\mu \nu}  - \frac{1}{2}g_{\mu \nu} T \Big)\chi^{\mu} n^{\nu} d V   \\ & = 2 \int_{\Sigma_t}\left(M_{\mu \nu} - \frac{1}{2}g_{\mu \nu} M  + E_{\mu \nu}\right) \chi^{\mu} n^{\nu} d V,
\end{aligned}
\end{equation}
where we used the decomposition $T_{\mu \nu}= M_{\mu \nu} +E_{\mu \nu}$, and $M_{\mu \nu}$ and $E_{\mu \nu}$ are given by Eqs.~\eqref{eq:fluid-emt} and \eqref{eq:max-emt}, respectively. 

Let us now examine each term on the right-hand side of Eq.~\eqref{eq:mJap}. The first term, which involves the matter energy-momentum tensor, contains the terms $M_{\mu \nu}\chi^\mu n^\nu$ and $- \frac{1}{2}g_{\mu \nu} M \chi^\mu n^\nu $. Using Eqs.~\eqref{eq:fluid-emt}, \eqref{eq:4velocity}, and \eqref{eq:chi}, we obtain 
\begin{align}
    M_{\mu \nu}\chi^\mu n^\nu & = - \rho_m {\cal F} u_\nu n^\nu, \\ 
    - \frac{1}{2}g_{\mu \nu} M \chi^\mu n^\nu &  = - \frac{1}{2}\left(-\rho_m +3p\right)\mathcal{F}u_\mu n^\mu,
\end{align}
respectively.
Adding the last two equations yields
\begin{align}
    \left( M_{\mu \nu} - \frac{1}{2}g_{\mu \nu} M\right)  \chi^\mu n^\nu    =- \frac{1}{2}\left(\rho_m +3p\right)\mathcal{F} u_\mu n^\mu. \label{eq:smarrM}
\end{align}
Consequently, integrating this expression over $\Sigma_t$ gives
\begin{align}
\int_{\Sigma_t}& \left(M_{\mu \nu} - \frac{1}{2}g_{\mu \nu} M \right) \chi^{\mu} n^{\nu} d V \nonumber \\ & = -\frac{1}{2}\int_{V_m}\left(\rho_m + 3p \right) \mathcal{F} u_{\mu} n^{\mu} d V. \label{eq:smarrsteps5}
\end{align}

The second term on the right-hand side of Eq.~\eqref{eq:mJap}, which  involves the electromagnetic energy-momentum tensor $E_{\mu\nu}$, contains the terms $ F_{\mu \alpha} F_{\nu}^{\ \alpha} \chi^{\mu}n^\nu$ and  $-\frac{1}{4} g_{\mu \nu}F_{\alpha \beta} F^{\alpha \beta} \chi^{\mu}n^\nu $.

Using the definition  $F_{\mu \nu} =  \nabla_{\nu} A_{\mu} - \nabla_{\mu} A_{\nu}$, and Eqs.~\eqref{eq:gaugepot} and~\eqref{eq:chi}, the first part of the electromagnetic EMT contracted with $ \chi^{\mu}$ furnishes 
\begin{equation}
\begin{aligned}
    F_{\mu \alpha} F_{\nu}^{\ \alpha} \chi^{\mu} & = F_{\nu}^{\ \alpha} \nabla_{\alpha}(\phi + \Omega \psi) - \psi  F_{\nu}^{\ \alpha} \nabla_{\alpha} \Omega \\
    &= \nabla^{\alpha}(F_{\nu \alpha} \Phi) - \Phi \nabla^{\alpha}F_{\nu \alpha} - \psi  F_{\nu}^{\ \alpha} \nabla_{\alpha} \Omega, \label{eq:smarrsteps1}
\end{aligned}
\end{equation}
where we used $F_{\mu \nu} \xi^{\mu} =  \nabla_{\nu} \phi, \, F_{\mu \nu} \eta^{\mu} = \nabla_{\nu} \psi$ in the first equality. To establish these identities, we used the fact that the electromagnetic gauge potential inherits the stationarity and axial symmetry of the spacetime. Consequently, $\xi^\nu\nabla_\nu\phi=\partial_t\phi=0$ and $\eta^\nu\nabla_\nu\phi=\partial_\varphi\phi=0$, with analogous relations holding for $\psi$.

Similarly, the second part gives, 
\begin{align}
& -\frac14 g_{\mu \nu}  F_{\alpha \beta} F^{\alpha \beta} \chi^{\mu}  = \frac{1}{2} \chi_{\nu} (\nabla_{\alpha}A_{\beta}) F^{\alpha \beta} \label{eq:smarrsteps2}\\ 
& = \frac{1}{2}\Big[\nabla_\alpha (A_\beta F^{\alpha \beta} \chi_\nu) - A_\beta F^{\alpha \beta} \nabla_\alpha \chi_\nu - A_\beta \chi_\nu \nabla_\alpha F^{\alpha \beta} \Big]. \nonumber
\end{align}
In turn, the second term in the last equality of Eq.~\eqref{eq:smarrsteps2} can be rewritten as
\begin{equation}
\begin{aligned}
   A_\beta & F^{\alpha \beta} \nabla_\alpha \chi_\nu  =
    A_\beta F^{\alpha \beta} \nabla_\alpha\left( \xi_\nu + \Omega \eta_\nu\right) 
   \\
   & =
   A_\beta F^{\alpha \beta} \left(\nabla_\alpha \xi_\nu + \Omega \nabla_\alpha \eta_\nu + \eta_{\nu} \nabla_\alpha \Omega\right) 
   \\ & = - \nabla_\alpha \left(\chi^\alpha A_\beta F^{\beta}_{\ \nu}  \right) + A_\beta F^{\alpha \beta} \eta_{\nu} \nabla_\alpha \Omega. \label{eq:smarrsteps3}
\end{aligned}
\end{equation}
In the last step we used the stationarity and axial symmetry of both the electromagnetic field tensor and the electromagnetic four-potential, together with the Killing equation, and also $\eta^{\nu} \nabla_\nu \Omega = \partial_{\varphi} \Omega = 0$.

Combining Eqs.~\eqref{eq:smarrsteps1},~\eqref{eq:smarrsteps2}, and~\eqref{eq:smarrsteps3}, we obtain
\begin{align}
E_{\mu \nu} \chi^{\mu} & =  - \frac{1}{4 \pi}\left(\Phi \nabla^{\alpha}F_{\nu \alpha} + \psi  F_{\nu}^{\ \alpha} \nabla_{\alpha} \Omega \right) \nonumber \\
& - \frac{1}{8 \pi}\left(A_\beta \chi_\nu \nabla_\alpha F^{\alpha \beta} + A_\beta F^{\alpha \beta} \eta_{\nu} \nabla_\alpha \Omega \right)  \label{eq:Esmarrstep2a}\\ & \ \ \ + \frac{1}{8\pi} \nabla_{\alpha} \left( 2\Phi F_{\nu}^{\ \alpha} +  A^\beta F^{\alpha}_{\ \beta} \chi_\nu - A^{\beta} F_{\nu \beta} \chi^{\alpha} \right). \nonumber
\end{align}
Now using the Maxwell equations \eqref{eq:Maxw} and the identity $A_\beta \chi^\beta= \Phi$, it follows 
\begin{align} 
E_{\mu \nu} \chi^{\mu} & =   - \frac{1}{2} \Phi\,\rho_e u_\nu \nonumber \\
& -\frac{1}{8\pi}\left[  2\psi  F_{\nu}^{\ \alpha} \nabla_{\alpha} \Omega + A_\beta F^{\alpha \beta} \eta_{\nu} \nabla_\alpha \Omega  \right]  \label{eq:Esmarrstep2b}\\ & + \frac{1}{8\pi} \nabla_{\alpha} \left[2\Phi F_{\nu}^{\ \alpha} +  A^\beta F^{\alpha}_{\ \beta} \chi_\nu - A^{\beta} F_{\nu \beta} \chi^{\alpha} \right]. \nonumber
\end{align}
Finally, the last integral term of Eq.~\eqref{eq:mJap} gives 
\begin{align}
& \int_{\Sigma_t}E_{\mu \nu} \chi^{\mu} n^{\nu} d V  =  
- \frac{1}{2}\int_{V_m}  \rho_e \Phi u_{\mu} n^{\mu} d V \nonumber \\ 
& 
- \frac{1}{8\pi }\int_{V_m} \left[  2\psi  F_{\nu}^{\ \alpha} \nabla_{\alpha} \Omega + A_\beta F^{\alpha \beta} \eta_{\nu} \nabla_\alpha \Omega  \right]  n^{\nu} d V  \label{eq:Esmarrstep2c}\\ & + \frac{1}{8\pi}\int_{\Sigma_t} \nabla_{\alpha} \left[2\Phi F_{\nu}^{\ \alpha} + A_\beta F^{\alpha \beta} \chi_\nu - A^{\beta} F_{\nu \beta} \chi^{\alpha} \right] n^{\nu} d V, \nonumber
\end{align}
where $V_m \subset \Sigma_t$ is the region containing charged matter. 

The integrand of the last integral term in Eq. ~\eqref{eq:Esmarrstep2c} corresponds to the divergence of a $2$-form. Using Gauss's theorem, we can rewrite the last integral as a surface term, with this surface being the boundary of $\Sigma_t$ at infinity. Assuming that $F_{\mu \nu}$ decays sufficiently rapidly at infinity on $\Sigma_t$ (otherwise, the spacetime under consideration cannot be asymptotically flat), the last integral in the equation above vanishes. Therefore, we can rewrite Eq.~\eqref{eq:Esmarrstep2c} as
\begin{align}
 &    \int_{\Sigma_t}E_{\mu \nu} \chi^{\mu} n^{\nu} d V  = - \frac{1}{2}\int_{V_m}  \rho_e \Phi u_{\mu} n^{\mu} d V  \label{eq:smarrsteps4a} \\
 & -\frac{1}{8\pi} \int_{V_m}\left[2 \psi F_{\mu \nu} \left(\nabla^{\nu} \Omega\right)+  A^{\sigma} F_{\nu \sigma} \left(\nabla^{\nu} \Omega\right) \eta_\mu\right]  n^{\mu} dV. \nonumber 
\end{align}

Therefore, combining Eqs.~\eqref{eq:smarrsteps5} and \eqref{eq:smarrsteps4a}, we obtain Eq.~\eqref{eq:generalsmarr}.

Notice that on the chosen hypersurface $\Sigma _{t}$ (defined by $t=\mathrm{constant}$), the axial Killing vector $\eta ^{\mu }$ is tangent to $\Sigma _{t}$ and therefore satisfies $\eta^\mu n_\mu=0$. Consequently, the term proportional to this contraction in Eqs.~\eqref{eq:smarrsteps4a} and~\eqref{eq:generalsmarr} vanishes identically, regardless of the QBH limit. We nevertheless choose to retain this term explicitly in Eq.~\eqref{eq:generalsmarr} to display the full expression obtained from the derivation.

\bibliographystyle{apsrev4-2}

\end{document}